\documentclass{article}
\usepackage[utf8]{inputenc}
\usepackage{rotating}
\usepackage{csquotes}
\usepackage{authblk}

\title{Spatial Model Selection and Uncertainty Quantification: Comparing Continuous and Discrete Wound Healing Models }

\date{\today}

\author[1]{John T. Nardini}

\author[1]{Jana L. Gevertz}

\affil[1]{Department of Mathematics and Statistics, The College of New Jersey, Ewing, NJ, 08628, USA. \\
nardinij@tcnj.edu}

\usepackage{makecell}
\usepackage[square,sort,comma,numbers]{natbib}
\usepackage{multirow}

\usepackage{graphicx}
\usepackage{amsmath}
\usepackage{geometry}
\usepackage{units}
\usepackage{amsfonts}
\usepackage{amssymb}
\usepackage{pifont}
\usepackage[title]{appendix}%
\usepackage{pdflscape}
\usepackage[dvipsnames,table]{xcolor}
\usepackage{hyperref}
\usepackage{cleveref}
\usepackage{enumitem}
\usepackage{bm}
\usepackage{hyperref}
\usepackage{verbatim}
\usepackage{ulem}
\usepackage[utf8]{inputenc}
\usepackage[T1]{fontenc}
\usepackage[ruled,vlined]{algorithm2e}
\usepackage{todonotes}
\usepackage{comment}
\usepackage{placeins}

\newcommand{\old}[1]{}

\newcommand{\p}{\bm{p}}
\newcommand{\ptrue}{\p^\text{true}}
\newcommand{\pselect}{\bm{\p}^\text{select}}
\newcommand{\phat}{\hat{\p}}
\newcommand{\ppull}{p_{pull}}
\renewcommand{\rm}{r_m}
\newcommand{\rp}{r_p}
\newcommand{\pmean}{\p^\text{mean}}
\newcommand{\pmap}{\p^\text{MAP}}
\newcommand{\pmeanhat}{\phat^\text{mean}}
\newcommand{\pmaphat}{\phat^\text{MAP}}

\newcommand{\MSEmin}{\text{MSE}^\text{min}}
\newcommand{\MSEthres}{\text{MSE}^\text{thres}}

\newcommand{\modelvec}{\bm{y}^\text{model}}
\newcommand{\datavec}{\bm{y}^\text{data}}
\newcommand{\model}{y^\text{model}}
\newcommand{\PDE}{y^\text{PDE}}
\newcommand{\nPDE}{\tilde{y}^\text{PDE}}
\newcommand{\ABM}{y^\text{ABM}}
\newcommand{\PDEvec}{\bm{y}^\text{PDE}}
\newcommand{\ABMvec}{\bm{y}^\text{ABM}}
\newcommand{\data}{y^\text{data}}
\newcommand{\tdata}{t^\text{data}}
\newcommand{\xdata}{x^\text{data}}
\newcommand{\tdatavec}{\bm{t}^\text{data}}
\newcommand{\xdatavec}{\bm{x}^\text{data}}

\newcommand{\Tau}{\mathcal{T}}
\newcommand{\dcdt}{\dfrac{\partial \PDE}{\partial t}}
\newcommand{\dt}{\Delta t}

\begin{document}

\maketitle
\begin{abstract}
    All data-driven modeling tasks (\emph{e.g.}, parameter estimation, uncertainty quantification, and data forecasting) require the selection of a mathematical model. An overlooked aspect of model selection is modality; for example, there are no guidelines on when to use a partial differential equation (PDE) model or an agent-based model (ABM) for spatial processes. To address this, we created a model selection pipeline that uses approximate Bayesian computations to perform parameter estimation, uncertainty quantification, and model selection (using both information criteria and out-of-sample forecasting). Applying the pipeline to artificial datasets (generated from ABMs) reveals that while both modalities yield comparable parameter estimation performance, the ABM estimates exhibit higher uncertainty, and the PDE models compute more than 1,000$\times$ faster. Surprisingly, the mean-field PDE is often selected over the true generative ABM model using both information criteria and data forecasting. Applying the pipeline to public wound healing data indicates that a PDE model with cell pulling and a time delay is the most appropriate model for this data, however, this model has high levels of parametric uncertainty. This methodology establishes a preliminary framework for selecting the appropriate modeling modality for spatial biological data.
\end{abstract}

\section{Introduction}

Mathematical models play an important role in understanding complex biological systems. The last decade has seen a shift in the way these  models are reconciled with biological data, particularly regarding uncertainty quantification (UQ) \cite{smith_uncertainty_2013}. In mathematical biology, inverse UQ is common, using techniques like Bayesian inference and practical identifiability analysis to constrain parameters against noisy or sparse data~\cite{WU2021111460}. Forward UQ is also increasingly used, with techniques such as global sensitivity analysis and virtual populations quantifying how parameter uncertainty propagates to model predictions \cite{bai_virtual_2022, brookes_saltelli_2015, craig_practical_2023}.

While powerful, both forms of UQ rely on the selection of a pre-determined model. However, it is well-established that model dynamics can be strongly influenced by the specific mathematical formulation of model terms~\cite{KOMAROVA2010530, Poleszczuk2018,Kutuva2023,oh2025mathematicalformschemotherapyradiotherapy}. Thus, the reliability of any UQ analysis is fundamentally constrained by the underlying model choice. Unlike physics, where models are often constrained by first principles, biology lacks such governing laws~\cite{KOMAROVA2010530}. This means mathematical biologists have no definitive standard for correctness, rendering the selection of model structure an important, yet often unquantified, source of predictive uncertainty.

In the absence of ``ground truth'' models, the selection of a model in mathematical biology commonly relies on manual observation, trial and error, and researcher intuition. When quantitative methods are used in model selection, researchers often seek the most ``parsimonious'' model as quantified using information criteria (IC).  While different IC formulations exist, each provides a way to balance the goodness-of-fit of the model to the data with model complexity, as quantified through the number of (fit) parameters~\cite{10.1093/bib/bbz016, warne_using_2019}. Using the number of parameters to quantify model complexity, IC cannot account for another important source of complexity: the structure of the model itself. For example, stochastic models are inherently more complex than deterministic ones, yet IC may favor them if they have fewer fit parameters.

This limitation is most pronounced when choosing between spatial modeling modalities. The choice to use a spatial model is usually dictated by data availability. However, once a spatial framework is deemed necessary, the modeler must choose between partial differential equation (PDE) models and agent based models (ABMs). PDEs offer analytical tractability and computational efficiency but lack the granular detail of ABMs, which simulate discrete agents interacting via rule-based mechanics~\cite{FORDVERSYPT2021100340,metzcar_review_2019}. It has been speculated that ABMs' heavy computational requirements prohibit modelers' ability to calibrate these models to data \cite{simpson_reliable_2022, jain_smore_2022, nardini_learning_2021}. However, recent advances in high-performance computing and parallelization call into question whether calibration remains a significant barrier \cite{l_rocha_multiscale_2024, breitwieser_high-performance_2023,chen_span_2024}.

\begin{figure}[ht!]
    \centering
    \includegraphics[width=0.95\linewidth]{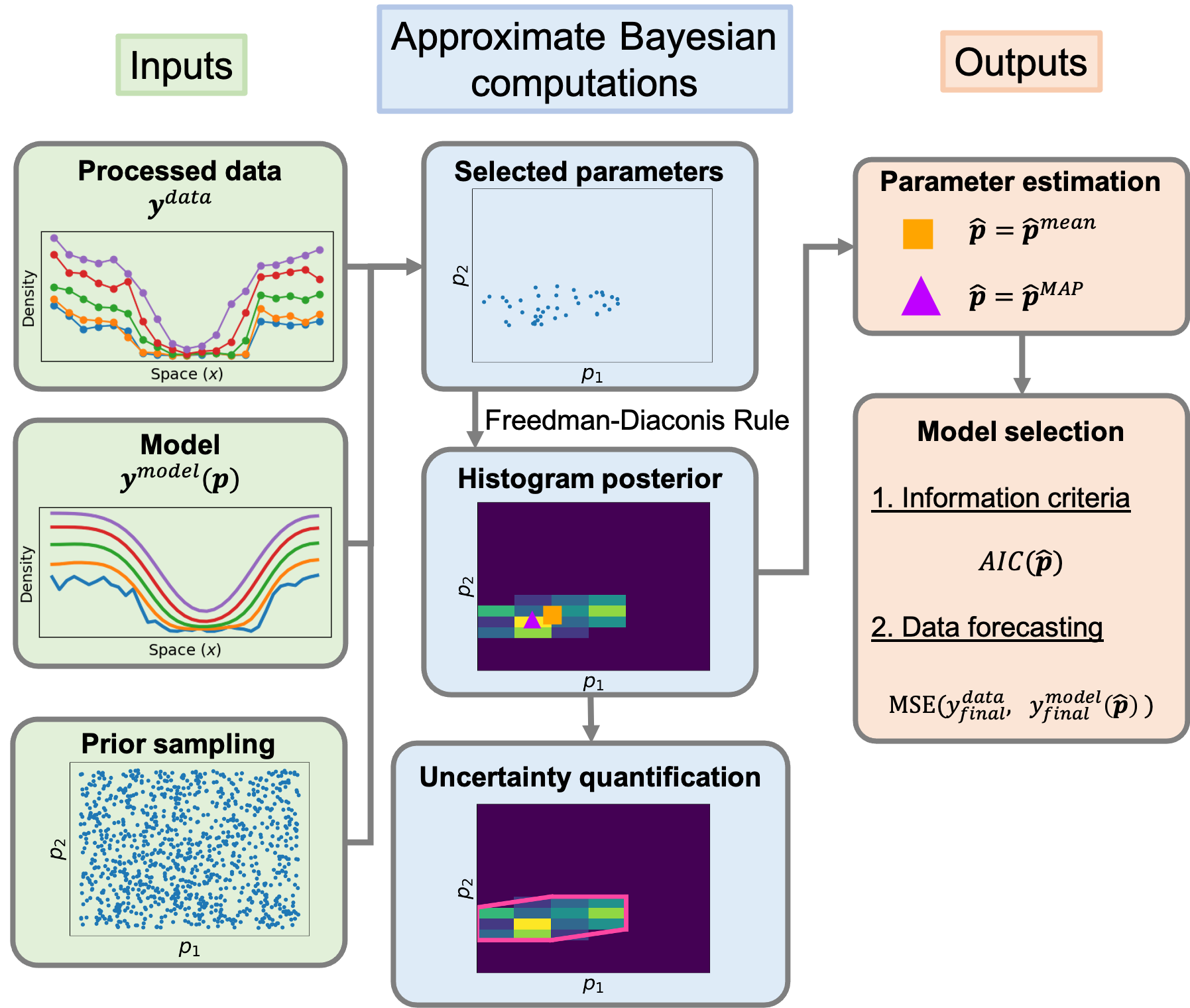}
    \caption{Model selection pipeline. Our ABC pipeline begins with the processed data, $\datavec$, and a prescribed model, $\modelvec(x,t; \p)$ with input parameters $\p\in\mathbb{R}^d$. We will consider both PDE and ABM models in our analysis. We sample $\p$ uniformly as a prior sampling. The ABC rejection method determines which prior samples are included in the selected set of parameters, $\pselect$.  We generate a $d$-dimensional histogram posterior from $\pselect$ using the Freedman-Diaconis Rule. The uncertainty of this posterior is quantified through a 90\% highest posterior density credible interval (magenta curve). From the histogram posterior, we estimate the best-fit parameter using the histogram mean, $\pmean$, or maximum a posteriori, $\pmap$. Either parameter estimate facilitates model selection using information criteria or out-of-sample forecasting. }
    \label{fig:PE_pipeline}
\end{figure}

The mathematical biology community currently lacks a systematic framework to select between distinct modeling modalities, such as PDE and ABM models. To address this, we present a pipeline for modality and model selection tailored to spatial biological data (Figure \ref{fig:PE_pipeline}). We validate this pipeline using synthetic datasets to assess parameter estimation and modality selection performance, and subsequently apply it to differentiate between candidate PDE and ABM models for wound healing data. The paper is organized as follows. We introduce the Artificial and Wound Healing datasets, candidate models of wound healing, and our model selection pipeline in Section \ref{sec:methods}. We then present our results when applying the model selection pipeline to the Artifical and Wound Healing datasets and comment on PDE and ABM model simulation timing in Section \ref{sec:results}. Section \ref{sec:conclusions} concludes by translating these findings into open questions regarding the signatures of datasets that necessitate agent based approaches. All code and simulated data to reproduce this work is publicly available at
\href{https://github.com/johnnardini/Spatial_model_selection_UQ/}{https://github.com/johnnardini/Spatial\_model\_selection\_UQ/}.

\section{Methods}\label{sec:methods}

\subsection{Description of the Wound Healing datasets} \label{sec:WH_data}

\begin{figure}[ht!]
    \centering
    \includegraphics[width=0.6\linewidth]{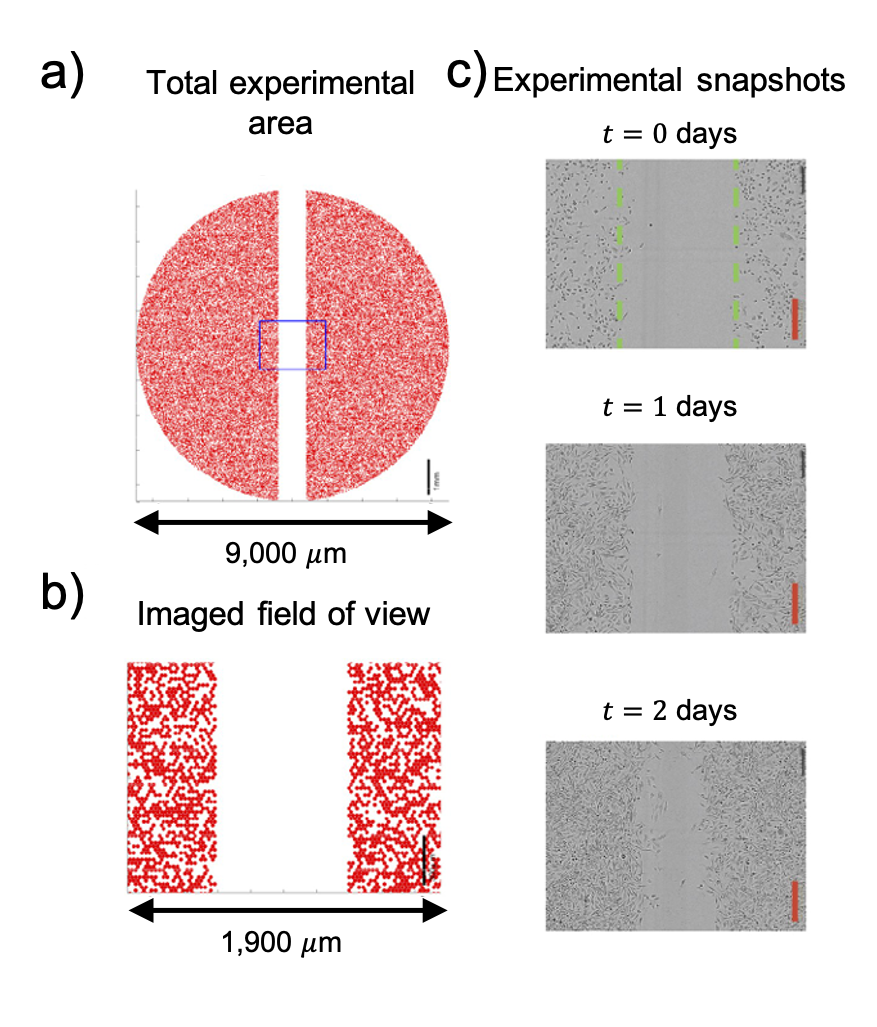}
    \caption{The scratch assay procedure (a) Schematic of the well population after the creation of an artificial wound. (b) Shows the field of view used for all images. (c) Example experimental snapshots after 0, 1, and 2 days. This figured has been reproduced and modified from \cite{lagergren_biologically-informed_2020}, Copyright Lagergren et al., and licensed under CC-BY 4.0 \href{https://creativecommons.org/licenses/by/4.0/}{https://creativecommons.org/licenses/by/4.0/}.}
    \label{fig:scratch_assay}
\end{figure}

We use publicly-available scratch assay data, originally published in \cite{jin_reproducibility_2016} and further analyzed in \cite{lagergren_biologically-informed_2020, vandenheuvel_computationally_2022}. To generate this data, cells from the PC-3 prostate cancer cell line were seeded in a well plate at six different densities: 10K, 12K, 14K, 16K, 18K, and 20K cells per well, where $n$K denotes $n\times1,000$. Wounds were artificially created in each experiment by manually removing cells from the middle portion of the well plate and washing off any excess cells from the denuded area (Figure \ref{fig:scratch_assay}(a)). The imaged field of view for each experiment is a rectangle 1900 $\mu$m wide and 1430 $\mu$m tall (Figure \ref{fig:scratch_assay}(b)). Images were recorded after 0, 0.5, 1, 1.5, and 2 days (Figure \ref{fig:scratch_assay}(c)). The experiment was performed in triplicate for all six densities.

\begin{figure}[ht!]
    \centering
    \includegraphics[width=0.99\linewidth]{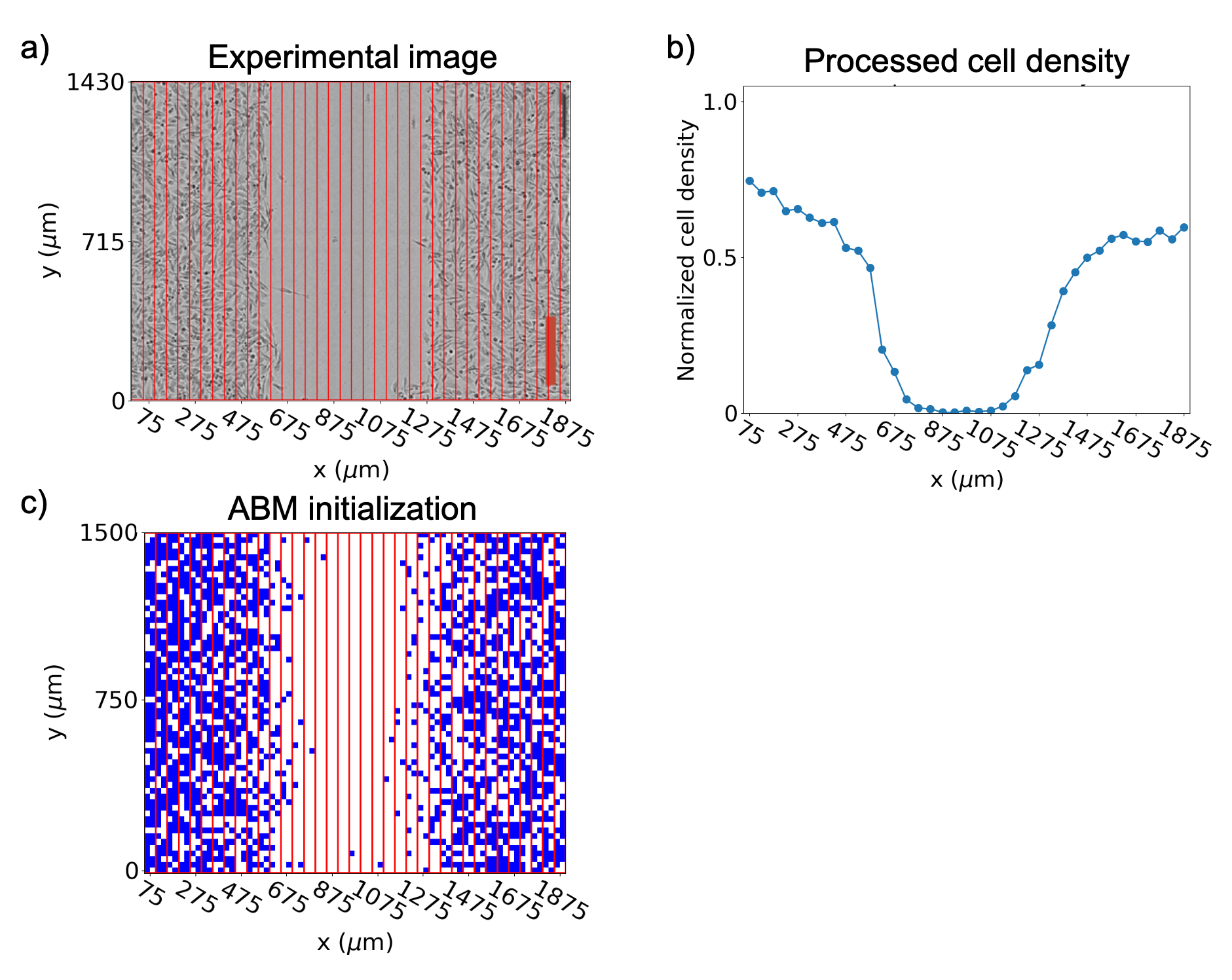}
    \caption{Estimating cell density from an experimental image and subsequent ABM initialization. (a) To quantify cell density in~\cite{jin_reproducibility_2016}, experimental images were split into 38 non-overlapping rectangles of width 50$\mu$m and height 1,430$\mu$m. (b) Each image was processed into a spatial cell density by counting the number of cells in each rectangle and dividing by the cell carrying capacity, $K$. The normalized cell densities are plotted against the horizontal center of each rectangle. (c) ABMs are initialized by populating agents in rectangles of width 50$\mu$m and height 1,500$\mu$m to match the normalized cell densities from panel (b).}
    \label{fig:cell_count_quantification}
\end{figure}

Each image was processed to estimate one-dimensional spatial density in \cite{jin_reproducibility_2016} by splitting the image into 38 non-overlapping and horizontally stacked rectangles of size 50$\times$1,430$\mu$m$^2$ (Figure \ref{fig:cell_count_quantification}(a)). The authors of \cite{jin_reproducibility_2016} then counted the number of cells inside each rectangle and divided by its area of $50\times1,430$= 71,500 $\mu$m$^2$. In this study, we compute a normalized density by dividing the number of cells in the rectangle by a carrying capacity, $K$; the value of $K$ is discussed in Section \ref{sec:ABM}. For all six cell density experimental conditions, we average the normalized densities across the three experimental replicates to obtain the final normalized density over space and time. Following \cite{lagergren_biologically-informed_2020}, we neglect the leftmost density value (corresponding to the rectangle centered at $x=25\mu m$) at all timepoints in each dataset because its values appear abnormally large. 

For notation, our spatial grid consists of the horizontal center of each rectangle, $\xdatavec=\{\xdata_i\}_{i=1}^{37}=\{75, 125, \dots, 1875, 1925 \ \mu\text{m}\}$, and the timepoints are given by $\tdatavec=\{t_j\}_{j=1}^5=\{0, 0.5, 1, 1.5, 2\ \text{hours}\}$. We write the averaged normalized density over space and time as
\begin{equation}
    \datavec = \{\data_{i,j}\}_{i=1,2,\dots,37}^{j=1,2,\dots,5}, \label{eq:y_data_WH}
\end{equation}
where $\data_{i,j}$ corresponds to the averaged normalized density value at location $\xdata_i$ and timepoint $\tdata_j$. For future out-of-sample forecasting (see Section \ref{sec:OOS_forecasting}), we let
\begin{equation}
    \datavec_\text{final} = \{\data_{i,5}\}_{i=1,2,\dots,37} \label{eq:y_data_WH_final}
\end{equation}
represent all spatial data at the final timepoint $t_\text{final}=t_5.$.

\subsection{Models of wound healing}

We develop two classes of models to represent the Wound Healing data from Section \ref{sec:WH_data}: PDE and ABM models. For each class, we will consider four specific models of increasing complexity. Each model includes a subset of the five unknown model parameters from Table \ref{tab:param_defns}, which can be estimated from data.

\begin{table}[ht!]
    \centering
    \begin{tabular}{|c|l|c|c|c|}
         \hline
         Parameter &  \multicolumn{1}{|c|}{Description} & \makecell{Range for \\ Artificial data} & \makecell{Range for \\ Wound Healing data} & Units \\
         \hline
         $\rm$ & Rate of agent migration  & $[0,\ 0.25]$ & $[0,\ 0.25]$ & $ \text{mm}^2$/day \\
         $\ppull$ & Pulling probability  & $[0,\ 1]$ & $[0,\ 1]$ & unitless \\
         $\rp$ & Rate of agent proliferation  & $[0,\ 3]$ & $[0,\ 5]$ &  1/day \\
         $a_0$ & Delay shift  & $[-20,\ 20]$ & $[-20,\ 20]$ & unitless \\
        $a_1$ & Delay steepness & $[0,\ 40]$ & $[0,\ 40]$ & 1/day \\
         \hline
    \end{tabular}
    \caption{Model parameter definitions. All PDE and ABM models considered require estimating a subset of the five presented parameters from data.}
    \label{tab:param_defns}
\end{table}

\subsubsection{ABMs}\label{sec:ABM}

We define four lattice-based cellular automaton ABMs to describe the Wound Healing data from Section \ref{sec:WH_data}. To match the images' field of view and obtain a similar carrying capacity, each ABM is defined on a two-dimensional rectangular lattice with $L_x=74$ horizontal lattice sites and $L_y=60$ vertical lattice sites. Each lattice site is square with side length $\Delta=25\ \mu m$ to represent the diameter of single cell from the PC-3 cell line \cite{park_morphological_2014}. An agent in our simulation corresponds to a cell in the scratch assay experiment. We assume agents can only occupy one lattice site, and that lattice sites can only be occupied by a single agent at a time. Reflecting boundary conditions are used at each boundary: any attempt to cross a boundary is aborted. All ABM models are simulated using the Gillespie algorithm \cite{Gillespie}.

Previous studies suggest that delayed reaction-diffusion equations are suitable to study this data \cite{lagergren_biologically-informed_2020,vandenheuvel_computationally_2022}. As such, we next define our model rules governing the delay in cell activity levels, migration, and proliferation. For notation throughout, $A_{i,j}(t)$ denotes that the $(i,j)^\text{th}$ lattice site is occupied at time $t$, and $0_{i,j}(t)$ denotes that the lattice site is empty. \vspace{.25cm}

\noindent \uline{Cell activity delay term}: Previous studies found that cell migration and proliferation for the Wound Healing datasets are delayed over time, with lower rates at early timepoints and larger rates later on \cite{lagergren_biologically-informed_2020,vandenheuvel_computationally_2022}. Following \cite{lagergren_biologically-informed_2020}, we use a sigmoidal function to capture this delay in cell activity in response to the wound. We define this delay as:
\begin{equation}
    \Tau(t)=\dfrac{1}{1+e^{-(a_1t+a_0)}}, \label{eq:tau}
\end{equation}
where $a_1\geq 0$ and $a_0$ are parameters that determine the timing and rate of the increase, and $0 \leq \Tau(t) \leq 1$. The parameter $a_0$ governs the horizontal translation of the curve; a smaller value shifts the inflection point to the right, effectively postponing the period of rapid increase. The parameter $a_1$ dictates the steepness of the rise, where larger values result in a sharper, more rapid progression toward saturation. \vspace{.25cm}

\noindent \uline{Migration rule}: Previous studies have reported that cells in the Wound Healing datasets are more migratory in high-density environments \cite{lagergren_biologically-informed_2020,vandenheuvel_computationally_2022}. One possible explanation for this behavior is that migratory cells use cell-cell adhesions to pull their followers with them \cite{chappelle_pulling_2019,nardini_forecasting_2024}. We incorporate this possible cell pulling behavior in our migration rule. 

We assume agents migrate with time-varying rate $R_m=\Tau(t)r_m$; once an agent elects to migrate, it chooses one of its four neighboring lattice sites to migrate into with equal probability. If the chosen lattice site is empty, then the agent shifts its location; otherwise, the migration event is aborted. If the migratory agent has a neighbor in the direction opposite of its migration direction, then it will pull that neighbor with it with probability $\ppull$. We summarize this process when the agent is moving rightwards as follows:
\begin{align}
    0_{i,j-1}(t)\ + \ A_{i,j}(t)\ +\ &0_{i,j+1}(t)\ \xrightarrow{\hspace{.85cm}R_m/4\hspace{.87cm}}\ &0_{i,j-1}(t+\dt)\ + \ 0_{i,j}(t+\dt)\ + \ A_{i,j+1}(t+\dt), \nonumber \\
    A_{i,j-1}(t)\ + \ A_{i,j}(t)\ +\ &0_{i,j+1}(t)\ \xrightarrow{\hspace{.56cm} \ppull R_m/4\hspace{.62cm}}\ &0_{i,j-1}(t+\dt)\ + \ A_{i,j}(t+\dt)\ + \ A_{i,j+1}(t+\dt), \nonumber \\
    A_{i,j-1}(t)\ + \ A_{i,j}(t)\ +\ &0_{i,j+1}(t)\ \xrightarrow{\hspace{.35cm} (1-\ppull) R_m/4\hspace{.25cm}}\ &A_{i,j-1}(t+\dt)\ + \ 0_{i,j}(t+\dt)\ + \ A_{i,j+1}(t+\dt), \tag{Migration rule} \label{eq:migration_rule_2}
\end{align}
with similar rules defined for agents migrating in the other three directions. In Equation \eqref{eq:migration_rule_2}, the top equation corresponds to a migrating agent with no neighbor, the second equation corresponds to a successful pulling event, and the third equation corresponds to an unsuccessful pulling event. All rates are divided by four because the agent migrates into one of its four neighboring sites with equal probability. \vspace{.25cm}

\noindent \uline{Proliferation rules}: In our cell proliferation rule, agents proliferate with time-varying rate $R_p=\Tau(t)r_p$; once an agent elects to proliferate, it chooses one of its four neighboring lattice sites to proliferate into with equal probability. If the chosen lattice site is empty, then the agent places its daughter agent there; otherwise, the proliferation event is aborted. We summarize this process when the agent is proliferating rightwards as follows:
\begin{align}
    A_{i,j}(t)\ +\ 0_{i,j+1}(t)\ \xrightarrow{\hspace{.25cm}R_p/4\hspace{.25cm}}\ A_{i,j}(t+\dt)\ + \ A_{i,j+1}(t+\dt). \tag{Proliferation rule} \label{eq:proliferation_rule}
\end{align}

\noindent \uline{Final ABMs considered}: The four ABMs we consider are presented in Table \ref{tab:AB_models_defn}. Each model is defined from assumptions on whether cell pulling occurs in \ref{eq:migration_rule_2}, and whether \ref{eq:migration_rule_2} and \ref{eq:proliferation_rule} are time delayed. The ``No pulling'' assumption is equivalent to setting $\ppull=0$ and the the ``No delay'' assumption is equivalent to $\Tau(t)=1$ for all time (which occurs as $a_0\rightarrow\infty$).

\begin{table}[ht!]
    \centering
    \begin{tabular}{|c|c|l|}
    \hline
          & Assumptions & \makecell{Unknown parameters}\\
         \hline
         ABM/PDE Model 1 &  \makecell{No delay ($\Tau(t)=1$) and \\No pulling ($\ppull=0$)}  & $\p=(\rm,\rp)^T$\\
         \hline
         
         ABM/PDE Model 2 &   No delay ($\Tau(t)=1$)  & $\p=(\rm,\ppull,\rp)^T$\\

         \hline
         ABM/PDE Model 3 &  No pulling ($\ppull=0$)  & $\p=(\rm,\rp,a_0,a_1)^T$\\

         \hline
         ABM/PDE Model 4 & None  & $\p=(\rm,\ppull,\rp,a_0,a_1)^T$\\
     \hline
         
    \end{tabular}
    \caption{ABMs and PDEs of collective migration and their corresponding 
    parameters. Each of the four ABMs consists of \ref{eq:migration_rule_2} and \ref{eq:proliferation_rule}, coupled with assumptions on whether cell pulling and the time delay occur. Each of the four PDEs consists of Equation \eqref{eq:PDE}, coupled with the same assumptions.}
    \label{tab:AB_models_defn}
\end{table} \vspace{.25cm}

\noindent \uline{Summarizing ABM output}. The experimental data is collected on the spatial grid $\xdatavec,$ which has a spacing of $50\ \mu$m between consecutive gridpoints, whereas the ABM has a spacing of $25\ \mu$m to reflect the diameter of a single cell. We compute the following ABM summaries for direct comparison between ABM simulations to the normalized experimental data. Let $A^{(r)}_{i,j}(t)$ represent the occupancy status of the $(i,j)^\text{th}$ lattice site over time from the $r^\text{th}$ of $R$ identically-prepared ABM simulations. 

We calculate the one-dimensional normalized density of the simulation at time $t$ by counting the number of agents in each consecutive pair of columns and dividing by $K=2L_y=120$, which represents the maximum number of agents that can occupy two columns: $$y^{ABM, (r)}_i(t)=\sum_{i'=1}^{L_y} \dfrac{A^{(r)}_{i',2i-1}(t) + A^{(r)}_{i', 2i}(t)}{K},\ \ \  i=1,2,\dots,L_x/2=37; \ \ r=1,2,\dots,R.$$ We record this normalized density at each  $t_j\in\tdatavec$ to match the timegrid of the data. We then compute the averaged normalized ABM density over all $R$ simulations as: $$\ABMvec = \{\langle \ABM_{i,j} \rangle \}_{i=1,2,\dots, 37}^{j=1,2,\dots,5} =  \left\{ \dfrac{1}{R} \sum_{r=1}^R y^{ABM, (r)}_i(t_j)\right\}_{i=1,2,\dots, 37}^{j=1,2,\dots,5}.$$ \vspace{.25cm}

\noindent \uline{Model initialization}: To initialize an ABM simulation for the Wound Healing datasets, we populate the ABM lattice based on the cell count data at the first time point (Figure \ref{fig:cell_count_quantification}(c)). Recall that $\data_{i,1}$ denotes the normalized cell density at the rectangle horizontally centered at $\xdata_i=25+50\times i\ \mu$m for $i=1,2,\dots,37$. Multiplying $\data_{i,1}$ by $K=120$ provides the number of cells in each rectangle; we then randomly populate this number of agents into the columns of the lattice that correspond to $\xdata_i$ (e.g., $\xdata_1$ corresponds to the first two columns, $\xdata_2$ corresponds to the next two columns, etc.).

\subsubsection{PDE models}

For comparison with the ABMs, we consider four PDE models. Each PDE Model is a reaction-diffusion model that results from coarse-graining the four ABMs into their PDE Model equivalents. Information on how to coarse-grain ABMs can be found elsewhere \cite{chappelle_pulling_2019,nardini_forecasting_2024}. Coarse-graining \ref{eq:migration_rule_2} and \ref{eq:proliferation_rule} together results in the following mean-field PDE model:
\begin{equation}
    \dcdt = \Tau(t) \left[ \dfrac{\partial}{\partial x}\left( \left( \dfrac{\rm}{4}+\ppull (\PDE)^2 \right) \dfrac{\partial \PDE}{\partial x}\right) + \rp \PDE\left( 1 - \dfrac{\PDE}{K} \right) \right], \label{eq:PDE}
\end{equation}
where $\PDE=\PDE(x,t)$ describes the number of ABM agents over space and time, and $K$ represents the population carrying capacity. Recall that in Section \ref{sec:ABM}, we used a carrying capacity of $K=120$ cells to occupy each $50\times1430\ \mu m^2$ rectangle. For consistency, we use this same carrying capacity for the PDEs in this study, but note that previous studies used a carrying capacity of $1.7\times10^{-3}$ cells/$\mu$m$^2$, which corresponds to about 121.5 cells in this same rectangular area \cite{jin_reproducibility_2016}.

The four PDE models we consider are presented in Table \ref{tab:AB_models_defn}. Similar to the ABM models, Each PDE is defined by assuming if cell migration and proliferation are delayed ($\Tau(t)=1$ if no delay) and if cell pulling occurs ($\ppull=0$ if no pulling occurs). To compare each PDE model simulation to the normalized cell density values, we normalize the PDE data by computing $\nPDE=\PDE/K$.

\subsection{Artificial data generation}

To verify our ability to estimate parameters and select the model underlying a noisy dataset, we construct eight Artificial datasets by simulating one of the ABMs at a given parameter vector $\p=\ptrue$. To match that the Wound Healing data is averaged over three replicates, we average this data over $R=3$ ABM simulations. Each Artificial dataset is summarized in Table \ref{tab:simulated_data_defn}. 

To initialize the Artificial datasets, we imitate the scratch assay process by populating the first and final 25\% of columns in the ABM at 20\% confluency. The remaining columns are all empty at initialization.

\begin{table}[ht!]
    \centering
    \begin{tabular}{|c|c|l|}
    \hline
        Dataset number & True ABM & \multicolumn{1}{|c|}{$\ptrue$}\\
    \hline
         1 & Model 1  & $\p=(\rm,\ \rp)^T=(0.03125,\ 0.5)^T$ \\
         2 & Model 1  & $\p=(\rm,\ \rp)^T=(0.03125,\ 2.0)^T$ \\
         3 & Model 1  & $\p=(\rm,\ \rp)^T=(0.15625,\ 0.5)^T$ \\
         4 & Model 1  & $\p=(\rm,\ \rp)^T=(0.15625,\ 2.0)^T$ \\
         5 & Model 2  & $\p=(\rm,\ \ppull,\ \rp)^T=(0.03125,\ 0.5,\ 0.5)^T$ \\
         6 & Model 3  & $\p=(\rm,\ \rp,\ a_0,\ a_1)^T=(0.15625,\ 2.5,\ 0,\ 5)^T$ \\
         7 & Model 3  & $\p=(\rm,\ \rp,\ a_0,\ a_1)^T=(0.15625,\ 2.5,\ -1.25,\ 5)^T$ \\
         8 & Model 4  & $\p=(\rm,\ \ppull,\ \rp,\ a_0,\ a_1)^T=(0.15625,\ 0.5,\ 2.5,\ 0,\ 5)^T$ \\
     \hline
    \end{tabular}
    \caption{Artificial dataset generation. The Artificial datasets are generated by the ABM with different underlying ABMs and parameter values.}
    \label{tab:simulated_data_defn}
\end{table}

\subsection{Parameter estimation}\label{sec:PE}

The first output of our pipeline is parameter estimation (Figure \ref{fig:PE_pipeline}). Here, we calibrate a parameterized mathematical model, $\modelvec(\p)$ for $\p\in\mathbb{R}^d$, to a noisy dataset, $\datavec$, by determining the parameter value estimate, $\hat{\p}$, that leads to close agreement between the two: $\modelvec(\hat{\p})\approx\datavec.$ To achieve this, we implement an Approximate Bayesian Computation (ABC) rejection pipeline \cite{vo_quantifying_2015}. The pipeline inputs a prior sampling for $\p$, from which it selects a sampling of parameters to compute the posterior distribution. From this distribution, we estimate the best-fit parameter, $\phat$ . We perform parameter estimation using both modeling approaches, e.g., $\modelvec(\p)=\ABMvec(\p)$ and $\modelvec(\p)=\PDEvec(\p)$. \vspace{.25cm}

\noindent \uline{Prior parameter sampling and selection}: To estimate $\p\in\mathbb{R}^d$, we first generate $10^4$ prior samples using a $d$-dimensional uniform distribution, with the bounds for each parameter provided in Table \ref{tab:param_defns}. For each parameter sample, $\p$, we compute the mean-squared error (MSE) between $\datavec$ and $\modelvec(\p)$:
\begin{equation}
    \text{MSE}(\p) = \sum_{i=1}^{37} \sum_{j=1}^5 (\data_{i,j} - \model_{i,j}(\p))^2.
\end{equation}
Let $\MSEmin$ denote the smallest observed MSE value from the $10^4$ samples.  We then determine the threshold MSE value for inclusion into the posterior sampling as 25\% higher than $\MSEmin$, e.g., $\MSEthres=1.25\times\MSEmin$ and $\bm{\p}^\text{select}=\{\p\ |\ \text{MSE}(\p) \le \MSEthres \}$. If this threshold results in fewer than 5 selected samples, we increase it by $0.25\times\MSEmin$ until at least 5 samples are included. \vspace{.25cm}

\noindent \uline{Posterior generation}: We generate a posterior distribution from $\bm{\p}^\text{select}$ by constructing a $d$-dimensional histogram. For each dimension $k\in\{1,2,\dots,d\}$, the bin width $h_k$ is determined independently using the Freedman-Diaconis Rule:
\begin{equation}
    h_k = \dfrac{2IQR(\pselect_k)}{\sqrt[3]{n^\text{select}}},
\end{equation}
where $IQR(\bm{x})$ represents the interquartile range (e.g., the difference between the $75^\text{th}$ and $25^\text{th}$ quartiles of $\bm{x}$), $\bm{\p}^\text{select}_k$ represents the $k^\text{th}$ component of all vectors in $\bm{\p}^\text{select}$, and $n^\text{select}$ represents the number of samples in $\bm{\p}^\text{select}$. This approach yields a grid of $B$ total hyperrectangular bins, each with volume $\prod_{k=1}^dh_k$. For each bin, we compute the the number of parameter samples inside the bin ($n_l$) and the center of the bin ($\bm{c}_l$) for $l=1,2,\dots,B.$ \vspace{.25cm}

\noindent \uline{Final parameter estimate determination}: From the posterior histograms, we generate a final output parameter estimate in two ways. First, we compute the sample mean of the histogram:
\begin{equation}
    \pmean = \dfrac{1}{B}\sum_{l=1}^B n_l\bm{c}_l.
\end{equation}
Second, we determine the bin center with the largest count of parameter samples to compute the maximum a posteriori (MAP) parameter estimator:
\begin{equation}
    \pmap = \{ \bm{c}_{l^*}\ |\ n_{l^*} \ge n_l \text{ for all } l=1,2,\dots,B \}.
\end{equation}

To compare a parameter estimate, $\phat$, to a known true parameter value, $\ptrue$, we report the relative Euclidean distance (RED) between the two:
\begin{equation}
    \text{RED}(\phat,\ptrue)=\sqrt{\sum_{i=1}^d \left( \dfrac{\hat{p}_i-p_i^\text{true}}{p_i^\text{true}} \right)^2}. \label{eq:RED}
\end{equation}

\subsection{Uncertainty quantification}\label{sec:methods_UQ}

We quantify the uncertainty associated with each parameter estimate by estimating a 90\% highest posterior density credible interval  (CI) for $\p$ (Figure \ref{fig:PE_pipeline}). To compute this interval from the histogram posterior, we normalize the bin counts $n_l$ of the posterior histogram to determine the probability density for each bin: $p_l=n_l/n^\text{select}$. The CI is then formed by iteratively including the bins (or hyperrectangles, in higher dimensions) with the highest probability values until the cumulative probability exceeds 0.9. This approach ensures the interval contains the most likely parameter values while excluding the 10\% of the total probability mass with the lowest density values. We visualize this multidimensional region by defining its boundary as the convex hull of all corners of the included hyperrectangles.

\subsection{Model selection using information criteria}

In this work, we fit eight total models (all four PDE and all four ABM models) to each dataset. 
To perform model selection, we determine which model minimizes the Akaike Information Criterion (AIC) for each dataset. For a model with best-fit parameter vector $\hat{\bm{\p}}$, the AIC evaluates the model's performance by balancing its goodness-of-fit (quantified via the MSE) against a penalty for model complexity, represented by the number of estimated parameters ($d$). 

Under the least squares framework, the AIC for each
model is calculated as \cite{malik_mathematical_2025,banks_aic_2017}:
\begin{equation}
    AIC = N(\ln(2\pi)+1) + N \ln(MSE(\hat{\bm{\p}})) + 2d, \label{eq:AIC}
\end{equation}
where $N=37\times5=185$ denotes the number datapoints in each dataset. The model with the lowest AIC score is considered the most parsimonious description of the data ensuring that additional complexity is only included in the selected model when it is justified by a sufficient improvement in fit. We compute the AIC for each model using either $\phat = \pmeanhat$ or $\phat = \pmaphat$.

\subsection{Model selection using out-of-sample forecasting}\label{sec:OOS_forecasting}

In addition to traditional model selection, we quantify each model's ability to predict unseen data. We implement data forecasting for each model by performing the ABC parameter estimation pipeline from Section \ref{sec:PE} on all spatial data from the first four timepoints, resulting in the parameter estimates $\pmeanhat$ and $\pmaphat$ for each model. We then simulate the model, initialized using the data at time 0, using either $\phat=\pmeanhat$ or $\phat=\pmeanhat$ to generate a model prediction for the final timepoint spatial data, given by $\modelvec_\text{final}(\phat)=\model(x,t_5,\phat)$. We report the MSE between the model prediction and true data at the final timepoint, $\datavec_\text{final}$. The model selected under this framework is the one that minimizes this MSE.

\section{Results and Discussion} \label{sec:results}

In this section, we apply our model selection pipeline to the Artificial datasets in Section \ref{sec:artifical_results} and the Wound Healing datasets in Section \ref{sec:WH_results}. We finish with a brief comparison between the computational times of PDE and ABM models.

Overall, we found better results using the mean parameter estimator, $\pmean$, over the MAP estimator, $\pmap$. For this reason, we present results using $\pmean$ in the main text, but refer readers to Appendix \ref{sec:MAP} for the results and discussion for $\pmap$.

\subsection{Artificial datasets}\label{sec:artifical_results}

For the Artificial datasets, we compare the PDE and ABM models' ability to estimate parameters (Section \ref{sec:artificial_PE}) and quantify uncertainty (Section \ref{sec:artificial_UQ}) before performing modeling selection using IC (Section \ref{sec:a_MS}) and out-of-sample forecasting (Section \ref{sec:a_OOS_forecast}).

\subsubsection{PDE and ABM models estimate parameters comparably well}\label{sec:artificial_PE}

We create the eight Artificial datasets from Table \ref{tab:simulated_data_defn} and run the ABC pipeline (Figure \ref{fig:PE_pipeline}) for both the PDE and ABM versions of the true underlying model (e.g., for Artificial datasets 1-4 we consider ABM Model 1 and PDE Model 1). We compare the PDE and ABM models' performance in estimating $\ptrue$ for Artificial datasets 1-4 (Table \ref{tab:artificial_PE_mean}). Both modeling approaches achieve comparable results, though the ABM's parameter estimates are more accurate for 3 of the 4 datasets. 

\begin{table}[ht!]
\begin{tabular}{|c|llll|}
\hline
 & \multicolumn{1}{c}{Dataset 1} & \multicolumn{1}{c}{Dataset 2} & \multicolumn{1}{c}{Dataset 3} & \multicolumn{1}{c|}{Dataset 4}\\
\hline
True & (0.031, 0.500)$^T$ & (0.031, 2.000)$^T$ & (0.156, 0.500)$^T$ & (0.156, 2.000)$^T$ \\
PDE 1 & (0.033, 0.488)$^T$ [0.065] & (0.034, 1.923)$^T$ [0.098] & (0.146, 0.501)$^T$ [0.065] & (0.169, 1.956)$^T$ [0.084] \\
ABM 1 & (0.036, 0.489)$^T$ [0.156] & (0.033, 1.998)$^T$ [0.071] & (0.150, 0.503)$^T$ [0.038] & (0.164, 2.001)$^T$ [0.051] \\
\hline
\end{tabular}
\caption{Parameter estimates for Artificial datasets 1-4 when using PDE and ABM Models 1 and the $\pmeanhat$ parameter estimate. We report the parameter estimate value and its RED value from Equation \eqref{eq:RED} in comparison to the true parameter, $\ptrue$. \label{tab:artificial_PE_mean}}
\end{table}

We next estimate $\ptrue$ from Artificial datasets 5-8 using the PDE and ABM versions of the true underlying model (e.g., Model 2 for Artificial dataset 5, Model 3 for Artificial dataset 6, etc. See Table \ref{tab:simulated_data_defn}). Both the PDE and ABM models obtain reasonable estimates for the $\rm$, $\ppull$, and $\rp$ parameters, however, neither model obtains accurate estimates for the $a_0$ and $a_1$ parameters (Table \ref{tab:artificial_PE2_mean}). 

\FloatBarrier 
\begin{sidewaystable}[htbp]
{\small
    \begin{tabular}{|c|llll|}
    \hline
  & \multicolumn{1}{c}{Dataset 5} & \multicolumn{1}{c}{Dataset 6} & \multicolumn{1}{c}{Dataset 7} & \multicolumn{1}{c|}{Dataset 8} \\
\hline
True & (0.031, 0.500, 0.500)$^T$ & (0.156, 2.500, 0.000, 5.000)$^T$ & (0.156, 2.500, -1.250, 5.000)$^T$ & (0.156, 0.500, 2.500, 0.000, 5.000)$^T$ \\
PDE & (0.033, 0.556, 0.475)$^T$, [0.138] & (0.166, 2.449, -4.167, 26.667)$^T$, [6.744] & (0.163, 2.386, -4.000, 20.000)$^T$, [3.721] & (0.172, 0.417, 2.325, -3.200, 25.833)$^T$, [5.920] \\
ABM & (0.034, 0.373, 0.505)$^T$, [0.271] & (0.151, 2.536, -1.778, 20.000)$^T$, [4.089] & (0.162, 2.317, -3.075, 22.960)$^T$, [3.878] & (0.152, 0.587, 2.512, 0.462, 28.615)$^T$, [4.757] \\
\hline
    \end{tabular}
    }
    \caption{Parameter estimates for Artificial datasets 5-8 when using  the PDE or ABM version of the true underlying model  and the $\pmeanhat$ parameter estimate. For Dataset 5, the true underlying model is ABM Model 2; for Datasets 6 and 7, the true underlying model is ABM Model 3; and for Dataset 8, the true underlying model is ABM Model 4. We report the parameter estimate value and its RED value from Equation \eqref{eq:RED} in comparison to the true parameter, $\ptrue$. \label{tab:artificial_PE2_mean}} 
\end{sidewaystable}
\FloatBarrier 

\subsubsection{ABM models lead to higher parameter uncertainty levels than PDE models}\label{sec:artificial_UQ}

\begin{figure}[ht!]
    \centering
    \includegraphics[width=0.5\linewidth]{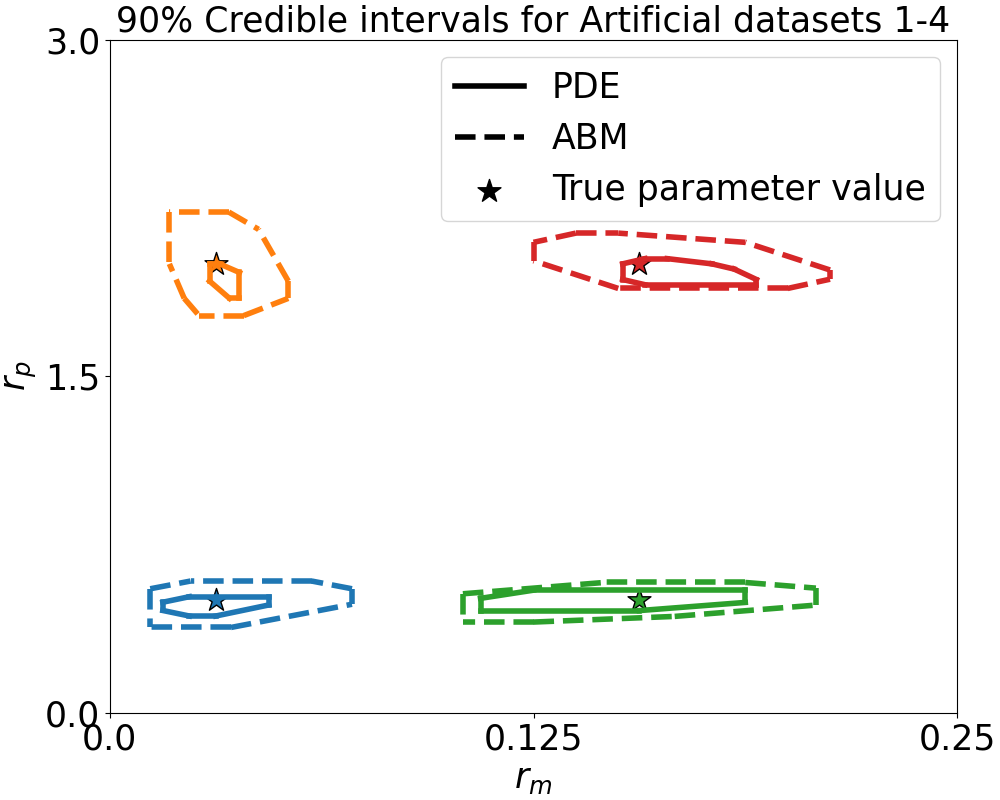}
    \caption{Uncertainty quantification for Artificial datasets 1-4. We compute 90\% credible intervals (see Section \ref{sec:methods_UQ}) for the Artificial datasets using either PDE (solid curves) or ABM (dashed curves) Model 1. The stars plot the true parameter values that generated each dataset. Results for Artificial dataset 1 are shown in blue, Artificial dataset 2 in orange, Artificial dataset 3 in green, and Artificial dataset 4 in red.}
    \label{fig:artificial_CI}
\end{figure}

Given our findings in Section \ref{sec:artificial_PE} that our ABC pipeline accurately estimates some parameters (namely, $\rm$, $\ppull,$ and $\rp$) while inaccurately estimating others ($a_0$ and $a_1$), we next aim to quantify the uncertainty associated with each parameter through the construction of 90\% credible interval estimates, as discussed in Section \ref{sec:methods_UQ}. We plot the 90\% credible intervals (CIs) associated with Artificial datasets 1-4 when using PDE and ABM Model 1 (Figure \ref{fig:artificial_CI}). Importantly, every CI contains the true underlying parameter value. However, the ABM produces wider intervals than the PDE, with the PDE CI always contained within the ABM CI. These results indicate that there is more uncertainty in estimating $\ptrue$ when using the ABM Model as compared to the PDE model. While this may initially seem surprising since the data are generated from the ABM, this discrepancy likely reflects the inherent stochasticity of the ABM framework in contrast to the deterministic nature of PDEs.

\begin{figure}[ht!]
    \centering
    \includegraphics[width=0.65\linewidth]{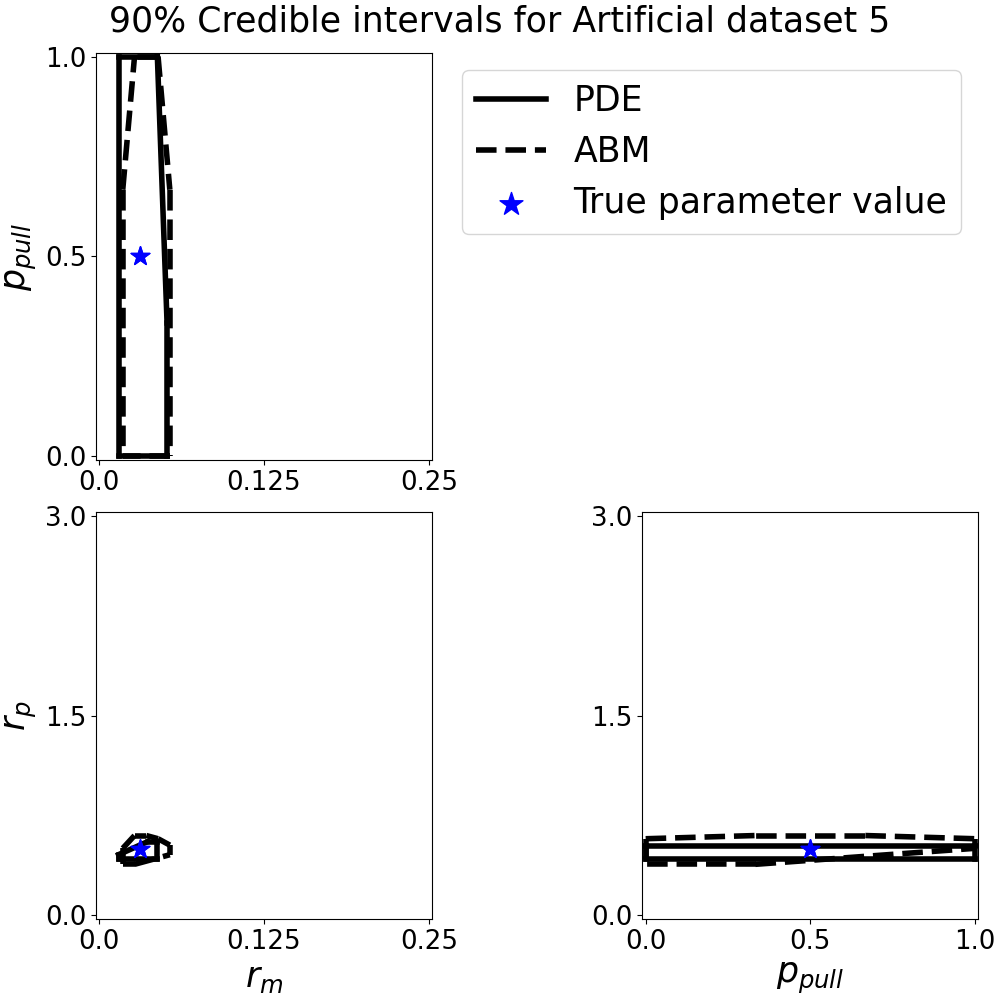}
    \caption{Uncertainty quantification for Artificial dataset 5. We compute marginal 90\% credible intervals (see Section \ref{sec:methods_UQ}) for each pair of parameter values for Artificial dataset 5 using using either PDE (solid curves) or ABM (dashed curves) Model 2. The stars plot the true parameter value pairs that generated the dataset.} 
    \label{fig:artificial_CI2}
\end{figure}

We plot the marginal 90\% CIs for each pair of parameters associated with Artificial dataset 5 that result when using either PDE or ABM Model 2 (Figure \ref{fig:artificial_CI2}). There is comparable uncertainty associated with each pair of parameters for both models. Notably, the marginal CIs for the $\rm$ and $\rp$ parameters are strict subsets of their considered parameter range (see Table \ref{tab:param_defns}), but the marginal CIs for $\ppull$ span the entire considered range. These $\ppull$ CIs suggest a lack of practical identifiability: there is insufficient data to confidently determine the value of this parameter.

We also plot the marginal 90\% CIs for Artificial datasets 6 (using both Model 3s), 7 (using both Model 3s), and 8 (using both Model 4s); see Supplementary Figures \ref{fig:artificial_CI6}, \ref{fig:artificial_CI7}, and \ref{fig:artificial_CI8}, respectively. Inspection of these shows wider CIs typically result when using ABM models as compared to PDE models, especially for Artificial datasets 7 and 8. The narrower PDE CIs usually still contain (or just barely miss) the true $\ptrue$ values, with the notable exception of the $a_1$ parameter for Artificial dataset 8. Similar to what was observed for Model 2 and Artificial dataset 5, the data are insufficient to confidently determine the value of at least one of the model parameters.

\subsubsection{Information criteria often select PDE models over ABMs}\label{sec:a_MS}

For the eight Artificial datasets from Table \ref{tab:simulated_data_defn}, we run the ABC parameter estimation pipeline for all eight models considered (four PDE and four ABM models). We report the AIC for each model using the $\pmeanhat$ estimate (Table \ref{tab:AIC_artificial_mean}): the most parsimonious model for each dataset is the model with the lowest AIC value. Notably, the true underlying ABM is never identified as the most parsimonious model by the AIC. This is likely because the inherent noise across stochastic realizations of the ABM leads to higher MSE values than those of its deterministic PDE counterparts, despite both being penalized equally for the same number of parameters. In contrast, the mean-field PDE approximation of the true ABM achieves the lowest AIC score in six of the eight datasets, and the second lowest in a seventh dataset. For Artificial dataset 7, the PDE approximation of the true ABM model ranks third by AIC score. In this instance, the model's superior fit is insufficient to offset the penalty imposed for having more parameters, leading the AIC to favor a simpler model structure.

\FloatBarrier 
\begin{sidewaystable}[htbp]
\begin{tabular}{|l|cccccccc|}
\hline
 & Dataset 1 & Dataset 2 & Dataset 3 & Dataset 4 & Dataset 5 & Dataset 6 & Dataset 7 & Dataset 8 \\
\hline
PDE Model 1 & \textbf{\cellcolor{green!25}220.0 (3.18)} & \textbf{\cellcolor{green!25}437.7 (10.31)} & \textbf{\cellcolor{green!25}285.7 (4.54)} & \textbf{\cellcolor{green!25}327.1 (5.67)} & \textbf{279.6 (4.39)} & 413.0 (9.02) & \uline{594.0 (24.01)} & 512.4 (15.45) \\
PDE Model 2 & \uline{226.6 (3.26)} & 438.9 (10.27) & \uline{292.5 (4.65)} & 342.5 (6.1) & \uline{\cellcolor{green!25}281.6 (4.39)} & 417.9 (9.17) & 604.7 (25.16) & 497.0 (14.06) \\
PDE Model 3 & 471.3 (12.11) & 447.0 (10.61) & 294.9 (4.66) & \uline{330.3 (5.65)} & 283.5 (4.39) & \textbf{\cellcolor{green!25}324.7 (5.48)} & \cellcolor{green!25}596.5 (23.82) & \uline{443.0 (10.38)} \\
PDE Model 4 & 228.7 (3.23) & 447.7 (10.54) & 296.4 (4.65) & 359.5 (6.54) & 291.8 (4.54) & 443.0 (10.27) & \textbf{414.6 (8.81)} & \textbf{\cellcolor{green!25}414.8 (8.82)} \\
ABM Model 1 & \cellcolor{blue!25}237.5 (3.5) & \uline{\cellcolor{blue!25}438.2 (10.34)} & \cellcolor{blue!25}299.8 (4.9) & \cellcolor{blue!25}332.9 (5.85) & 296.8 (4.82) & \uline{403.6 (8.58)} & 596.8 (24.37) & 507.9 (15.08) \\
ABM Model 2 & 250.8 (3.71) & 448.7 (10.83) & 315.0 (5.26) & 354.4 (6.5) & \cellcolor{blue!25}283.4 (4.43) & 419.9 (9.27) & 609.7 (25.86) & 506.3 (14.78) \\
ABM Model 3 & 330.7 (5.66) & 476.9 (12.48) & 318.7 (5.3) & 331.8 (5.69) & 297.6 (4.73) & \cellcolor{blue!25}434.6 (9.93) & \cellcolor{blue!25}659.0 (33.39) & 658.5 (33.29) \\
ABM Model 4 & 299.2 (4.72) & 500.3 (14.0) & 310.7 (5.03) & 418.8 (9.02) & 301.0 (4.77) & 543.4 (17.68) & 736.9 (50.33) & \cellcolor{blue!25}692.8 (39.64) \\
\hline
\end{tabular}
\caption{Model selection for the Artificial datasets using the $\pmeanhat$ parameter estimate. AIC scores (with MSE values in parentheses) for each Artificial dataset. For each column, \textbf{the bolded value indicates the model with the lowest AIC value}, and \uline{the underlined value indicates the model with the second-lowest AIC value}, the blue cell the true ABM model, and the green cell indicates the mean-field PDE counterpart of the true ABM model.} \label{tab:AIC_artificial_mean}
\end{sidewaystable}
\FloatBarrier

\subsubsection{Out-of-sample forecasting also selects PDE models over ABMs}\label{sec:a_OOS_forecast}

For the Artificial datasets, the AIC often identifies the mean-field PDE approximation as the most parsimonious model, rather than the ABM used to generate the data. Given that the Artificial datasets are generated from an ABM, we originally hypothesized that the true ABM structure would be favored in the selection process.  To investigate this further, we used an alternative model selection criterion based on out-of-sample forecasting: models were calibrated using only the first four time points, and their performance was evaluated by their ability to predict the final (fifth) time point. Under this approach, the ``best'' model is defined as the one achieving the minimum MSE at the held-out time point.

\begin{figure}[ht!]
    \centering
    \includegraphics[width=0.85\linewidth]{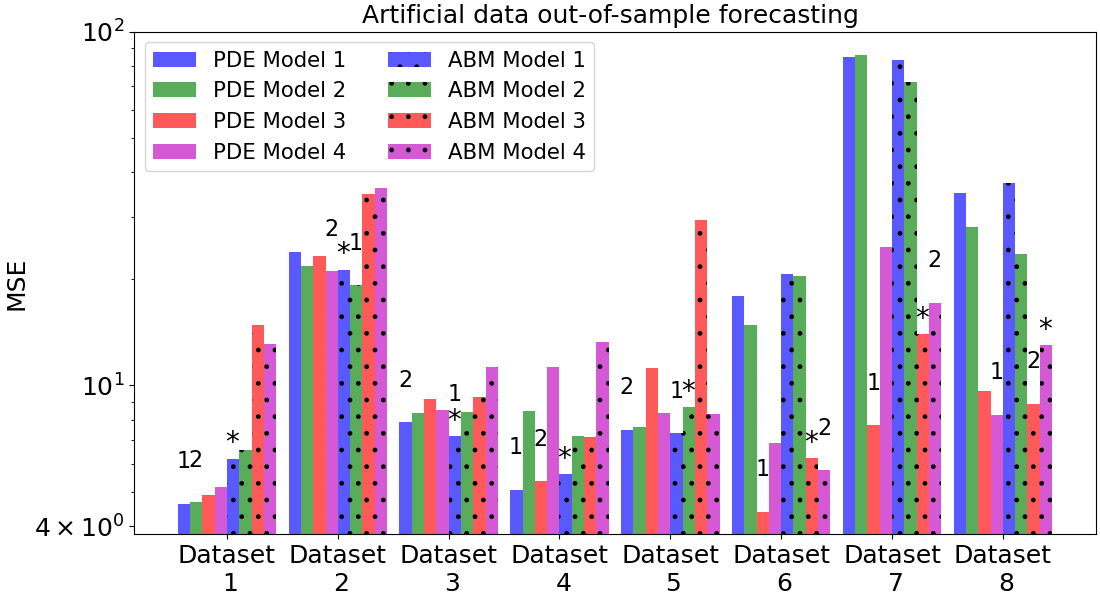}
    \caption{Out-of-sample forecasting for the Artificial datasets using the $\pmeanhat$ estimate. We depict the MSE between the final Artificial data timepoint, and the prediction of each model at the final timepoint using the $\pmeanhat$ parameter estimate (determined from all timepoints, excluding the final one). The true model used to generate each simulation is designated by the asterisk; the `1' and `2' designate the models with  best and second-best out-of-sample MSEs, respectively.}
    \label{fig:forecasting_artificial_mean}
\end{figure}

Specifically, we applied the ABC parameter estimation pipeline to all eight models, for all eight Artificial datasets, using the truncated data (time points 1–4). After simulating each model with the resulting best-fit parameter estimate, $\pmeanhat$, we calculated the MSE between the model's prediction and the observed data at the final time point (Figure \ref{fig:forecasting_artificial_mean}). Even under this predictive framework, the true ABM obtained the lowest MSE for only one dataset (Artificial dataset 3), whereas its mean-field PDE counterpart achieved the best performance in five of the eight datasets (Artificial datasets 1, 4, 6, 7, and 8).

It is instructive to compare the model selected by both AIC and out-of-sample forecasting with the true ABM model used to generate the Artificial datasets. For Artificial Artificial datasets 1, 4, 6, and 8, both approaches select the mean-field PDE approximation of the true ABM. For Artificial dataset 3, out-of-sample forecasting actually identifies the true ABM model as its first choice, with the corresponding PDE model as the second choice. This PDE model is the one selected by AIC. The remaining three datasets show more disagreement across model-selection approaches. Out-of-sample forecasting fails to identify the true ABM model (or its mean-field PDE counterpart), for both Artificial dataset 2 and 5. AIC successfully identifies both the true PDE and ABM models for Artificial dataset 2, and it identifies the true PDE as the second most parsimonious model for Artificial dataset 5. For Artificial dataset 7, AIC fails to identify the true model, whereas out-of-sample forecasting identifies the true ABM and its mean-field PDE as the top two models.

\subsection{Wound Healing data}\label{sec:WH_results}

For the Wound Healing datasets, we perform model selection using both information criteria (Section \ref{sec:WH_MS}) and out-of-sample forecasting (Section \ref{sec:WH_DF}) before investigating the parameter estimation and uncertainty quantification results (Section \ref{sec:WH_UQ}).

\subsubsection{IC selects PDE Model 4 for the Wound Healing data} \label{sec:WH_MS}

For the six Wound Healing datasets, we run the ABC parameter estimation pipeline for all eight considered models (four PDE and four ABM models). We report the AIC for each model using the $\pmeanhat$ estimate (Table \ref{tab:AIC_WH_mean}). For five of the six densities, PDE Model 4 achieves both the lowest MSE and AIC values, suggesting it is the model that best, and most parsimoniously, describes these data. The success of this model suggests that there is both cell pulling and a time delay present in the majority of the Wound Healing datasets considered, which is confirmed by the parameter estimates (Supplementary Table \ref{tab:parameter_estimation_WH}) and estimated delay terms, $\Tau(t)$, (Supplementary Figure \ref{fig:delay_terms_WH_PDE_model_4}). The remaining dataset was generated from the lowest density (10K cells per plate), and ABM Model 1 achieves the lowest AIC (and MSE) after calibration to this data, which would suggest there is neither cell pulling nor a time delay in the data. The lack of a delay in this data is further demonstrated through the estimated $\Tau$ term for this data from PDE Model 4, which reaches one (indicating the delay has completed) much sooner than $\Tau(t)$ for other datasets (Supplementary Figure \ref{fig:delay_terms_WH_PDE_model_4}).

\begin{table}[ht!]
\begin{tabular}{|l|cccccc|}
\hline
 & Density 10K & Density 12K & Density 14K & Density 16K & Density 18K & Density 20K \\
\hline
PDE Model 1 & 717.8 (46.87) & 660.3 (34.35) & 682.4 (38.72) & 684.2 (39.1) & 709.5 (44.81) & 773.5 (63.36) \\
PDE Model 2 & 716.7 (46.09) & 661.1 (34.13) & 659.4 (33.83) & 667.7 (35.38) & 705.7 (43.43) & 773.6 (62.7) \\
PDE Model 3 & 722.1 (46.95) & \uline{572.4 (20.91)} & \uline{657.3 (33.08)} & 601.7 (24.5) & 725.0 (47.69) & \uline{643.6 (30.71)} \\
PDE Model 4 & 842.4 (89.0) & \textbf{547.1 (18.04)} & \textbf{603.1 (24.41)} & \textbf{573.8 (20.84)} & \textbf{574.8 (20.95)} & \textbf{639.7 (29.76)} \\
ABM Model 1 & \textbf{700.8 (42.75)} & 650.1 (32.51) & 679.3 (38.07) & 678.5 (37.92) & 709.0 (44.69) & 767.9 (61.46) \\
ABM Model 2 & \uline{709.4 (44.32)} & 659.4 (33.82) & 662.1 (34.32) & 663.7 (34.61) & 707.2 (43.78) & 772.0 (62.16) \\
ABM Model 3 & 744.7 (53.06) & 978.1 (187.36) & 725.4 (47.8) & 755.3 (56.19) & 639.9 (30.12) & 651.0 (31.98) \\
ABM Model 4 & 739.5 (51.04) & 583.4 (21.95) & 710.4 (43.6) & \uline{600.4 (24.06)} & \uline{578.6 (21.39)} & 660.9 (33.37) \\
\hline
\end{tabular}
\caption{Model selection for the Wound Healing data using the $\pmean$ parameter estimator. AIC scores (with MSE values in parentheses) for each Artificial dataset. For each column, \textbf{the bolded value is the lowest AIC value}, and \uline{the underlined value is the second-lowest AIC value}.} \label{tab:AIC_WH_mean}
\end{table}

When calibrated to the Density 10K dataset, we find that  PDE Model 4 yields a higher MSE than the three simpler PDE variants. This result is counterintuitive, as these models are nested simplifications of PDE Model 4 and should, theoretically, achieve an inferior or equal fit. Further inspection (Supplementary Table \ref{tab:ABC_convergence_WH}) reveals that this discrepancy is an artifact of the ABC algorithm's convergence rate. Upon increasing the prior sample size to $N=10^6$, PDE Model 4 achieves the expected lower MSE. However, this marginal improvement in fit is insufficient to offset the AIC penalty for added complexity. As a result, simpler PDE models remain the more parsimonious choice (AIC of 700.8 versus 720) for the Density 10K dataset.

\subsubsection{Out-of-sample forecasting selects PDE Models 3 and 4 for the Wound Healing data} \label{sec:WH_DF}

\begin{figure}[ht!]
    \centering
    \includegraphics[width=0.85\linewidth]{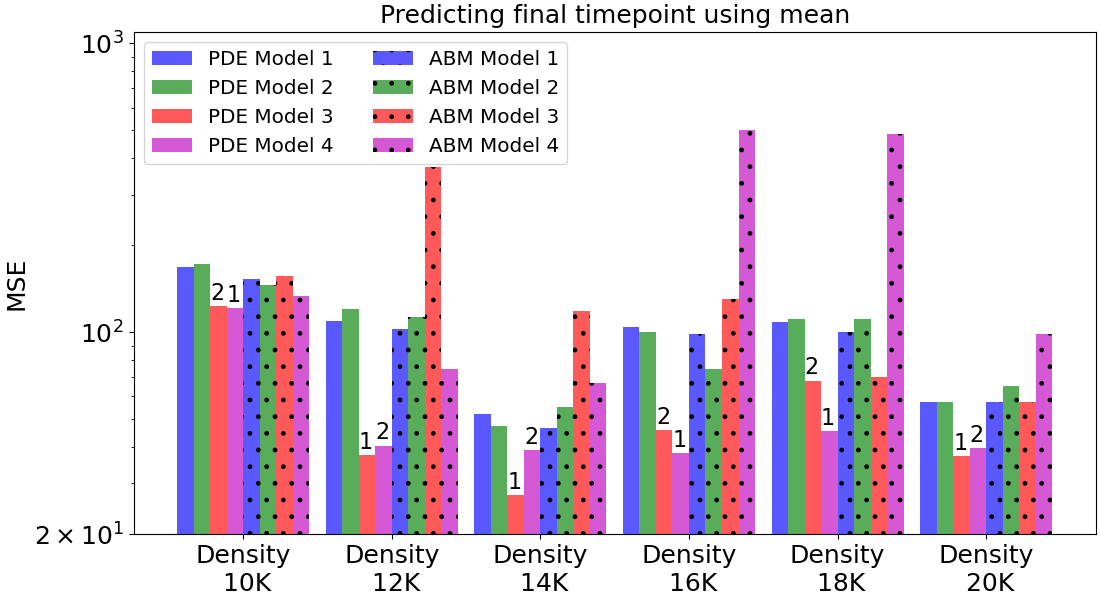}
    \caption{Out-of-sample forecasting for the Wound Healing datasets using the $\pmeanhat$ estimate. We depict the MSE between the final Wound Healing data timepoint, and the prediction of each model at the final timepoint using the $\pmeanhat$ parameter estimate (determined from all timepoints, excluding the final one). The `1' and `2' designate the models with  best and second-best out-of-sample MSEs, respectively.}
    \label{fig:forecasting_WH_mean}
\end{figure}

As with the Artificial datasets, we next used an alternative model selection criterion based on out-of-sample forecasting. The results for the six Wound Healing datasets, across the eight models, are shown in Figure \ref{fig:forecasting_WH_mean}. Consistent with our model selection results using the AIC, PDE Models 3 and 4 are the top performers for densities of 12K cells per plate and higher. To detail, PDE Model 3 achieves the best prediction for three of the six datasets (Densities 12K, 14K, 20K). For each of these densities, PDE Model 4 is the second most predictive model.

PDE Model 4 achieves the best prediction for the other three densities (10K, 16K, 18K). For densities of 16K and 18K cells per plate, this PDE model was also selected as the most parsimonious using AIC. Thus, there is consistency between our model selection results when using either AIC and out-of-sample forecasting. Overall, the results strongly indicate that a PDE model with a time delay (either with or without pulling), should be ``selected'' for the Wound Healing datasets, which is consistent with previous studies \cite{lagergren_biologically-informed_2020, vandenheuvel_computationally_2022}. The one exception is for the 10K density; when performing model selection using AIC, ABM Models 1 and 2 (without time delays) are selected. We consider the ABM model selection here an outlier in our data analysis results.

\subsubsection{PDE Model 4 has high levels of parametric uncertainty}\label{sec:WH_UQ}

\begin{figure}
    \centering
    \includegraphics[width=0.65\linewidth]{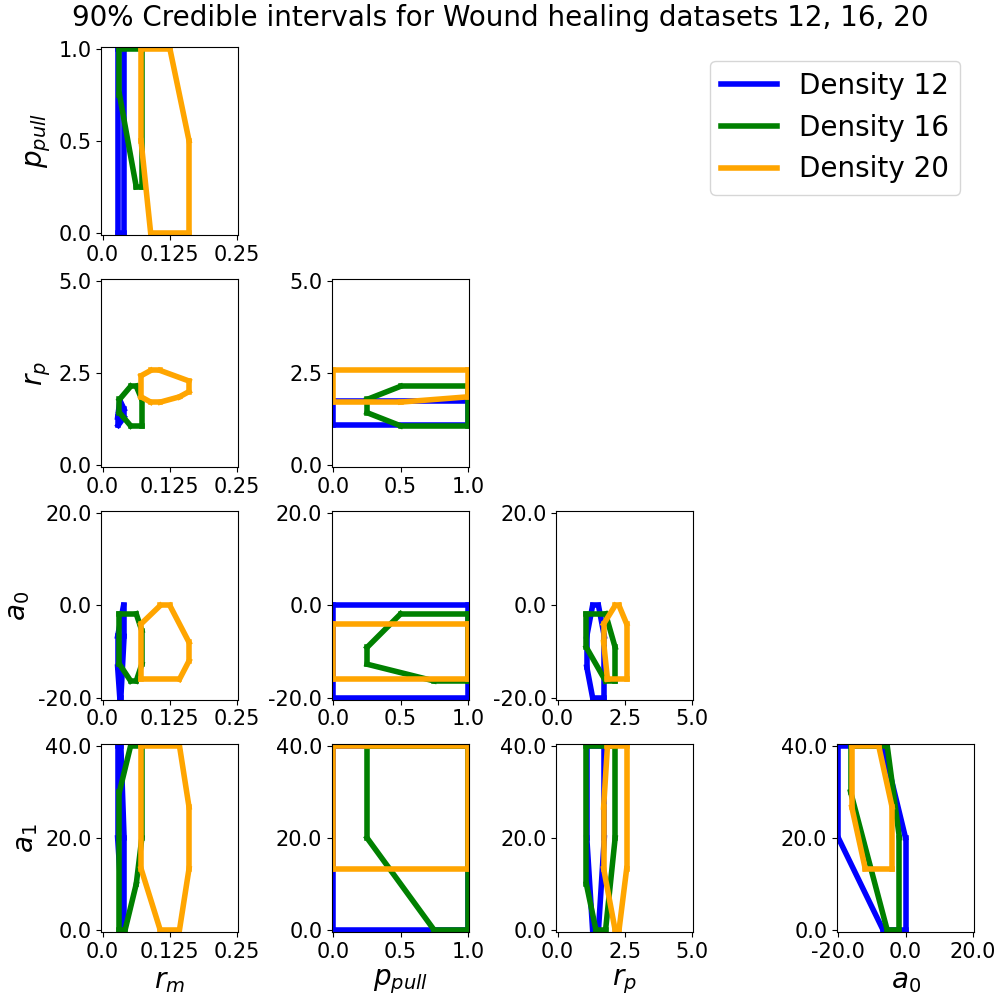}
    \caption{Uncertainty quantification for the the Wound Healing datasets. We compute 90\% credible intervals (see Section \ref{sec:methods_UQ}) for the Density 12K (blue), 16K (green), and 20K (orange) Wound Healing datasets using PDE Model 4. Results for Density 10K, 14K, and 18K not shown.}
    \label{fig:WH_CI}
\end{figure}

Sections \ref{sec:WH_MS} and \ref{sec:WH_DF} suggest that PDE Model 4 most parsimoniously describes the Wound Healing datasets and can be used to future data not used during the ABC pipeline. We now quantify the uncertainty associated with this model's parameter estimates for the Wound Healing data by plotting the 90\% credible intervals (CIs) associated with the Density 12K, 16K, and 20K datasets that result when using PDE Model 4 (Figure \ref{fig:WH_CI}; similar results obtained for the other densities, results not shown). Similar to our findings in Section \ref{sec:artificial_UQ}, we find that the CIs are most compact for the $\rm$ and $\rp$ parameters, indicating less uncertainty related to these parameters. As the density increases, we observe an increase in the values of $\rm$ and $\rp$ included in the CIs, suggesting the cells migrate and proliferate more at higher density.

The CIs for the $\ppull$ and $a_1$ parameters often span their entire prior ranges, indicating we cannot estimate these parameters from the data. Similarly, the CIs for $a_0$ often cover half of the parameter’s range; while this allows us to conclude that $a_0 < 0$, the specific value remains poorly constrained. Consequently, although both out-of-sample forecasting and the AIC favor PDE Model 4, several of its parameters cannot be uniquely determined. Experimentally measuring one or more of these parameters would likely reduce model uncertainty and enhance its overall predictive capability.

\subsection{PDE model simulations are more than 1,000$\times$ faster than the ABM models}\label{sec:timing}

\begin{figure}
    \centering
    \includegraphics[width=0.5\linewidth]{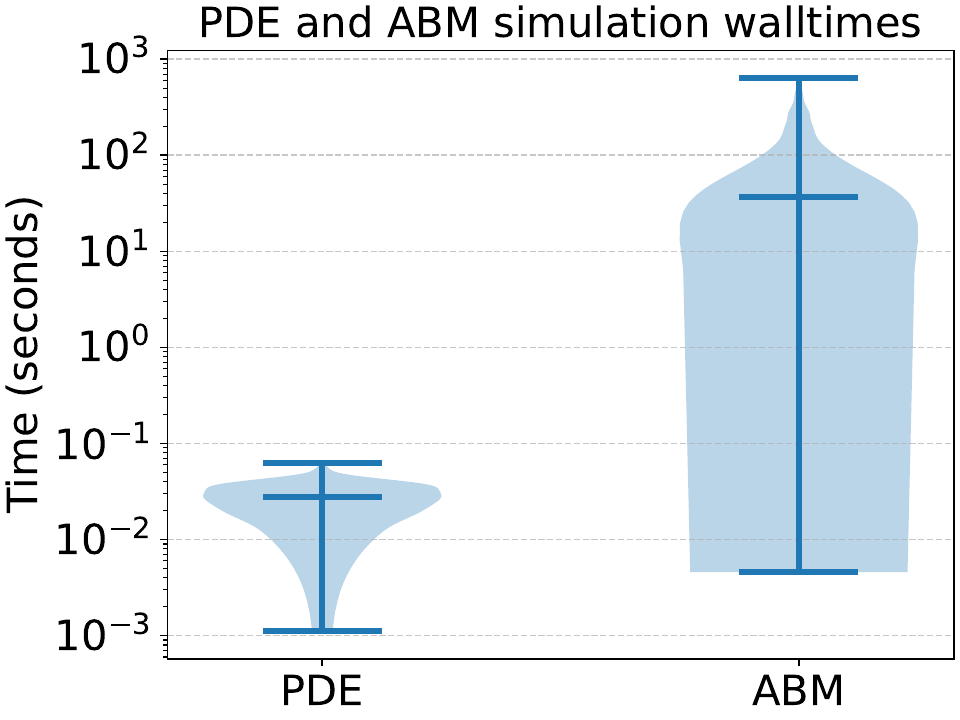}
    \caption{Walltimes for 1,000 PDE and ABM model simulations. The shaded areas of the violin plot represent the entire data range, with the bottom, middle, and top horizontal lines designating the minimum, median, and maximum values, respectively.}
    \label{fig:walltimes}
\end{figure}

To measure the differences in computation times between PDE and ABM model simulations, we performed 1,000 simulations from both models and recorded the walltimes of each simulation (Figure \ref{fig:walltimes}). Specifically, we simulated PDE and ABM Model 4 starting with the initial condition used for all Artificial datasets. We generated 1,000 samples for $\p$ uniformly from the parameter ranges outlined in Table \ref{tab:param_defns} for the Artificial datasets. The median PDE and ABM simulation walltimes were 0.028 and 37.0134 seconds, respectively, suggesting that PDE simulations are, more than 1,000 times faster to compute than ABM simulations.

\section{Conclusions and Future Work}\label{sec:conclusions}

When working with spatial biological data, PDE models are traditionally favored over ABM models for their computational efficiency \cite{baker_modeling_2026, kutz_data-driven_2025, nardini_learning_2021, warne_using_2019}. However, recent advances in high-performance computing have reduced the computational bottlenecks that hinder data-driven ABM modeling \cite{l_rocha_multiscale_2024, breitwieser_high-performance_2023,chen_span_2024}. In this study, we designed an ABC pipeline to determine how the choice of using a PDE or ABM framework impacts common data-driven modeling tasks, including parameter estimation, model selection, and uncertainty quantification. 

Motivated by publicly available data from scratch assay experiments, we considered models that describe cells' migratory and proliferative behaviors \cite{jin_reproducibility_2016}. Following \cite{lagergren_biologically-informed_2020,vandenheuvel_computationally_2022}, our models incorporate (1) cell pulling mechanisms that increase cells' migration rates and (2) a delay term quantifying how cells' migratory and proliferative responses to the wound change with time. When implementing our pipeline on Artificial datasets generated from the ABM models, we found that PDE and ABM models recovered the true underlying parameters with comparable accuracy levels, but the ABM model estimates were associated with increased levels of uncertainty (quantified using 90\% CIs). Surprisingly, our ABC pipeline often selected the mean-field PDE counterpart of the true ABM model that generated each Artificial dataset, as quantified by either the AIC or out-of-sample forecasting. PDE models with a time delay, and with or without cell pulling, were selected when applying our pipeline to the Wound Healing datasets. In addition to the PDE models' strong performance, we estimated their simulations are more than 1,000 time faster to compute than ABM models (Figure \ref{fig:walltimes}).

We designed our pipeline using an ABC approach due to its wide usage with ABMs \cite{vo_quantifying_2015, lambert_bayesian_2018, schalte_efficient_2020, duswald_calibration_2024}, however, there are known limitations to these algorithms \cite{beaumont_approximate_2019, pesonen_abc_2023}. In particular, they converge slowly with the number of prior samples, especially in higher dimensions. We computed $10^4$ prior samples per model to ensure consistency between models and to maintain a reasonable pipeline walltime for the ABMs. Models with more parameters (e.g., PDE and ABM Models 3 and 4) likely require additional prior samples to ensure convergence. In future work, we plan to extend our pipeline with more prior samples. This will be straightforward for the PDE models, which complete the current pipeline within seconds. For the ABM models, we will investigate two approaches: (1) efficient sampling methods and (2) surrogate modeling. For efficient sampling, we will compare the performance of several Markov-chain Monte Carlo approaches that have been been proposed for use with ABMs \cite{warne_multilevel_2018, beigmohammadi_trajectory-matching_2025, warne_multifidelity_2022}. For surrogate modeling, there has been much recent interest in the use of DEs, artificial neural networks, and/or Gaussian processes as computationally-efficient models to predict ABM behavior throughout parameter space \cite{angione_using_2022, van_der_hoog_surrogate_2019, comlekoglu_surrogate_2025, bergman_connecting_2024, jain_smore_2022, norton_advances_2026}. We will train each surrogate model on numerous ABM simulations to find which minimizes pipeline walltime while maintaining high fidelity to the ABM outputs. It is possible that with more prior samples (using either efficient sampling or surrogate modeling), the ABM models' performance will improve, for example through more accurate parameter estimates, less uncertainty related to the parameters, and/or better model selection results.

More prior samples may reduce parameter uncertainty, however, this will not occur if the model parameters lack identifiability \cite{chis_structural_2011}. This is of particular relevance to PDE Model 4, which was often selected as the best model to describe the Wound Healing datasets, yet it had several parameters with wide 90\% CIs (see Figure \ref{fig:WH_CI}). This was also seen in the models with time delay terms when applied to the Artificial datasets (see Supplementary Figures \ref{fig:artificial_CI6} - \ref{fig:artificial_CI8}), strongly suggesting that models with a time delay contain parameters that are not practically identifiable. In future work, we plan to precisely determine if the models' parameters are practically identifiable using profile likelihoods \cite{murphy_profile_2000,eisenberg_confidence_2017}. If the model parameters are not identifiable, we will determine if different data collection procedures (e.g., measuring cell behaviors and/or the delay rate prior to the experiment) can enhance parameter identifiability \cite{raue_structural_2009, raue_identifiability_2010, steiert_experimental_2012, munoz-tamayo_be_2018, wieland_structural_2021, gevertz_minimally_2024}. 

Both PDE and ABM models are widely used in the mathematical biology literature to describe spatial datasets \cite{nardini_learning_2021, nardini_forecasting_2024, chappelle_pulling_2019, turner_discrete_2004, simpson_reliable_2022}. PDE models provide a macroscale population summary (here, how the cell population responds to a wound), whereas ABM models provide a microscale lens by emulating individual agent trajectories (here, cells' response to the wound). Despite this additional information, we found no benefit in using ABM models over PDE models for performing parameter estimation, uncertainty quantification, and model selection for the Artificial and Wound Healing datasets in this study. 

However, there are likely many scenarios where the additional complexity of an ABM model is necessary to describe biological data. The ``rule of thumb'' in the literature is that ABM models are suited for describing more complex phenomena than PDE models \cite{an_optimization_2017, metzcar_review_2019, deangelis_decision-making_2019, glen_agent-based_2019}, but we lack clear guidelines on this choice. In the future, we plan to compare PDE and ABM models for various biological scenarios of increasing complexity (e.g., heterogeneous populations \cite{gallaher_spatial_2018}, self-organizing patterns \cite{dorsogna_self-propelled_2006, volkening_linking_2020, volkening_methods_2024}, and/or bistable behaviors \cite{colon_bifurcation_2015}) as first steps towards providing these guidelines for the community. Towards this end, we challenge the mathematical biology community to consider what the ``signature(s)'' of a biological dataset requiring an ABM model is. For example, what patterns are associated with datasets whose complexity, heterogeneity, and/or self-organizing behavior require the use of ABM modeling? How can modelers detect these patterns \emph{a priori} instead of directly comparing the results of both modeling frameworks?

\vspace{.5in}

\noindent\textbf{Declaration of competing interests:} Given their role on the editorial board of \emph{Mathematical Biosciences} and Guest Editor for the special issue ``Data-driven modeling for agent-based dynamics in biology'', John T. Nardini had no involvement in the peer-review of this article and has no access to information regarding its peer-review. Full responsibility for the editorial process for this article was delegated to another journal editor. Jana L. Gevertz declares no conflict of interest.\vspace{.125in}

\noindent\textbf{Data Availability statement:} All code and simulated data for this work is publicly available at\\
\href{https://github.com/johnnardini/Spatial_model_selection_UQ/}{https://github.com/johnnardini/Spatial\_model\_selection\_UQ/}.\vspace{.125in}

\noindent\textbf{Funding Statement:} The authors acknowledge use of the ELSA high performance computing cluster at The College of New Jersey for conducting the research reported in this paper. This cluster is funded in part by the National Science Foundation under grant numbers OAC-1826915 and OAC-2320244.\vspace{.125in}

\noindent\textbf{Declaration of generative AI and AI-assisted technologies in the manuscript preparation process:} During the preparation of this work the authors used OpenAI's ChatGPT in order to prepare the python code to generate, and sample from, the posterior histogram distributions. The authors used Google's Gemini to revise the manuscript text and generate documentation strings for all python code. After using these tools, the authors reviewed and edited the content as needed and take full responsibility for the content of the published article.

\bibliographystyle{unsrt}
\bibliography{references.bib}  
\newpage

\appendix
\renewcommand{\thesection}{S\arabic{section}}
\renewcommand{\thetable}{S\arabic{table}}
\renewcommand{\thefigure}{S\arabic{figure}}
\renewcommand{\theequation}{S\arabic{equation}}

\noindent {\Large{\textbf{Supplementary Information} }}
\section{Supplementary Figures}

\begin{figure}[ht!]
    \centering
    \includegraphics[width=0.7\linewidth]{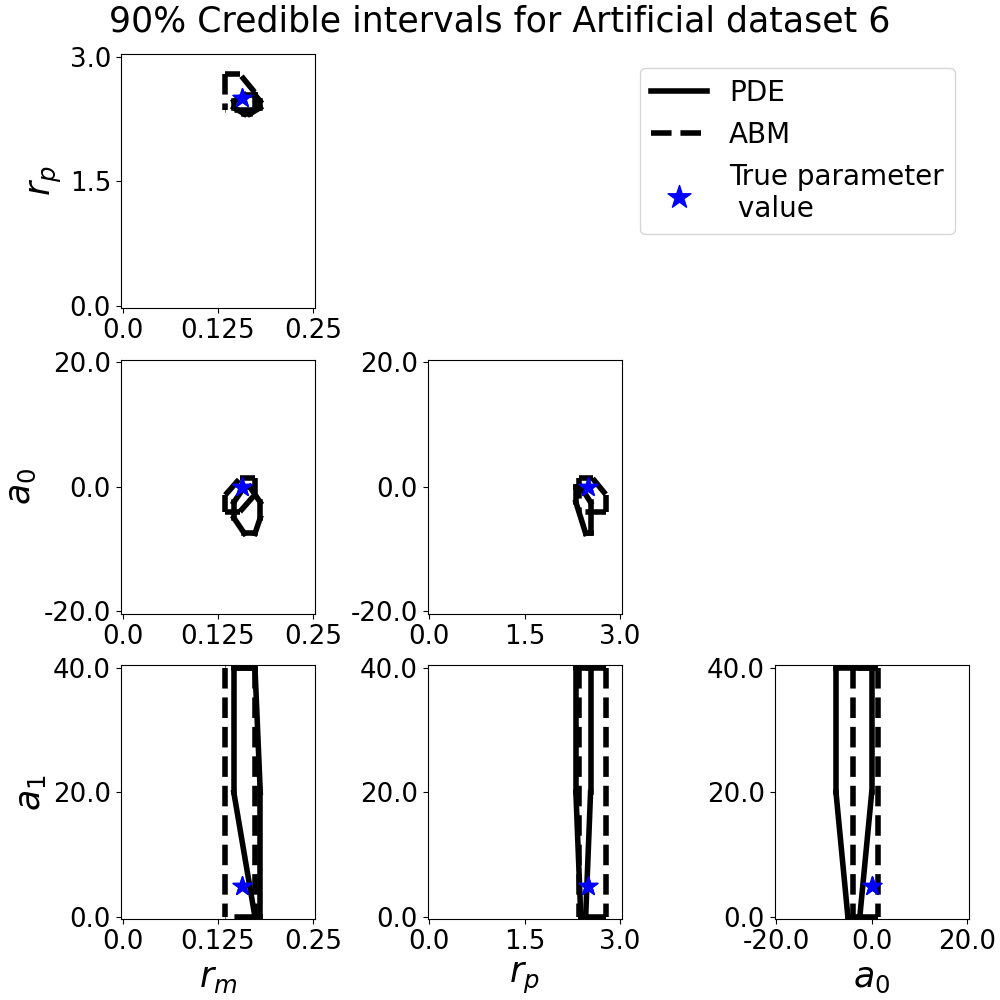}
    \caption{Uncertainty quantification for Artificial dataset 6. We compute 90\% credible intervals (see Section \ref{sec:methods_UQ}) for Artificial datasets using PDE (solid curves) or ABM (dashed curves) model 3. The stars plot the true parameter pairs that generated the dataset. }
    \label{fig:artificial_CI6}
\end{figure}

\begin{figure}[ht!]
    \centering
    \includegraphics[width=0.7\linewidth]{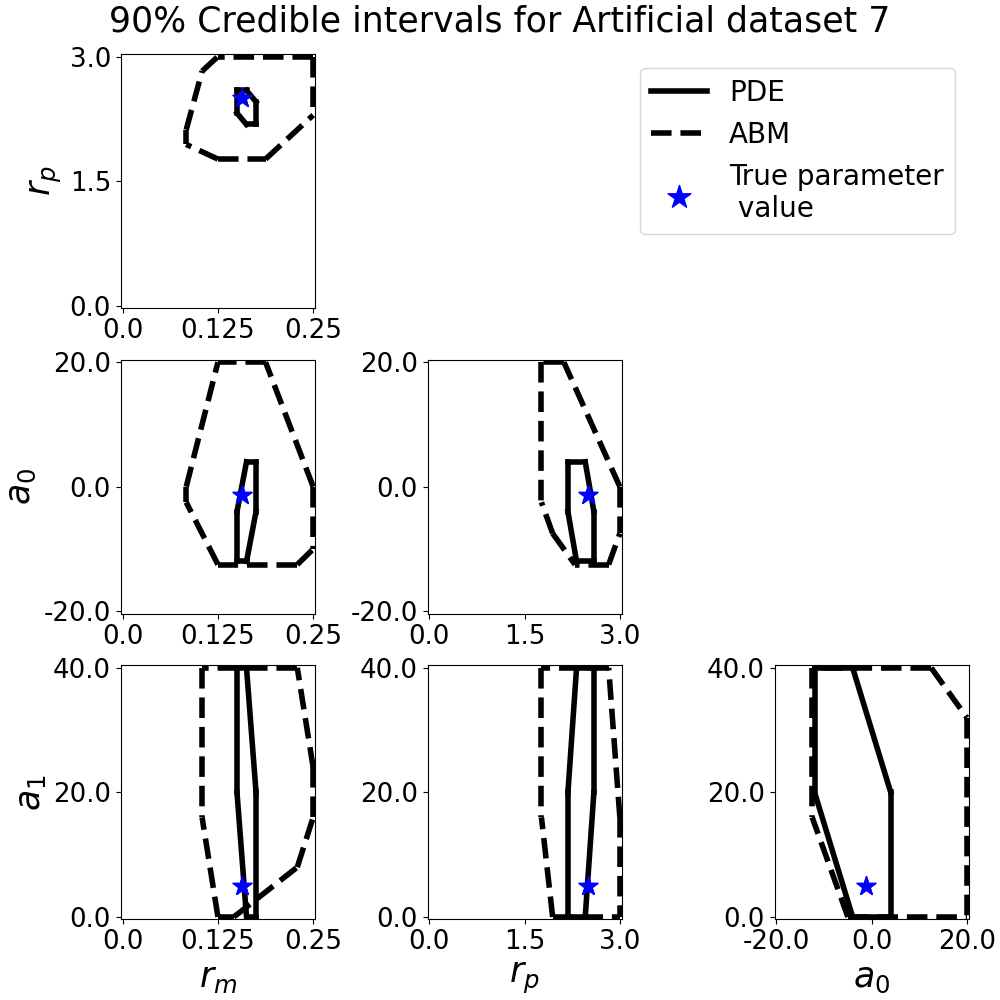}
    \caption{Uncertainty quantification for Artificial dataset 7. We compute 90\% credible intervals (see Section \ref{sec:methods_UQ}) for Artificial datasets using PDE (solid curves) or ABM (dashed curves) model 3. The stars plot the true parameter pairs that generated the dataset. }
    \label{fig:artificial_CI7}
\end{figure}

\begin{figure}[ht!]
    \centering
    \includegraphics[width=0.7\linewidth]{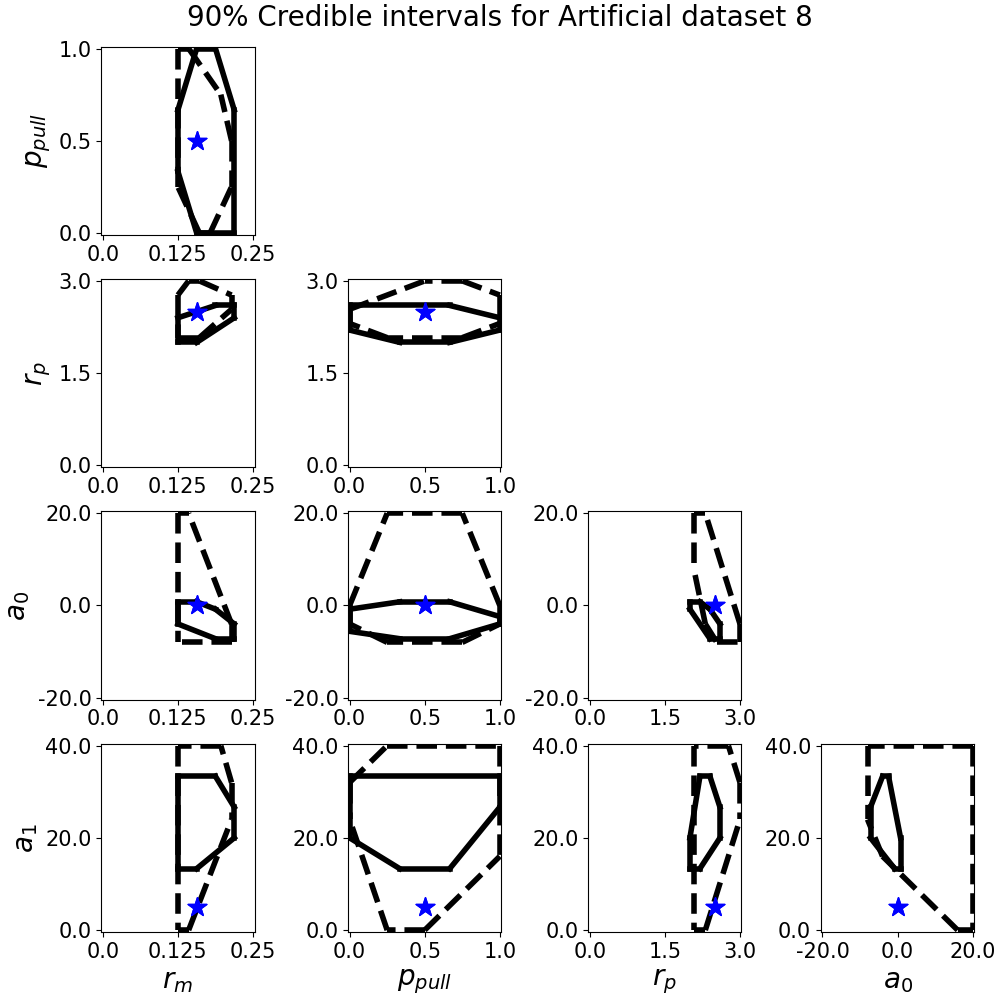}
    \caption{Uncertainty quantification for Artificial dataset 8. We compute 90\% credible intervals (see Section \ref{sec:methods_UQ}) for Artificial datasets using PDE (solid curves) or ABM (dashed curves) model 4. The stars plot the true parameter pairs that generated the dataset. }
    \label{fig:artificial_CI8}
\end{figure}

\begin{table}[ht!]
    \centering
    \begin{tabular}{|c|l|}
    \hline
    Dataset & \multicolumn{1}{|c|}{$\phat=(R_m,\ p_{pull},\ R_p,\ a_0,\ a_1)^T$} \\
    \hline
    Density 10K & (0.033, 0.672, 1.321, -0.276, 19.080)$^T$ \\
    Density 12K & (0.033, 0.500, 1.457, -11.333, 26.000)$^T$ \\
    Density 14K & (0.072, 0.607, 1.460, -7.033, 21.905)$^T$ \\
    Density 16K & (0.057, 0.681, 1.607, -8.889, 25.000)$^T$ \\
    Density 18K & (0.072, 0.500, 1.786, -10.000, 25.000)$^T$ \\
    Density 20K & (0.114, 0.467, 2.127, -10.174, 26.377)$^T$ \\
    \hline
    \end{tabular}
    \caption{Parameter estimates for the Wound Healing datasets when using PDE Model 4 and the $\pmeanhat$ parameter estimate.}
    \label{tab:parameter_estimation_WH}
\end{table}

\begin{figure}
    \centering
    \includegraphics[width=0.5\linewidth]{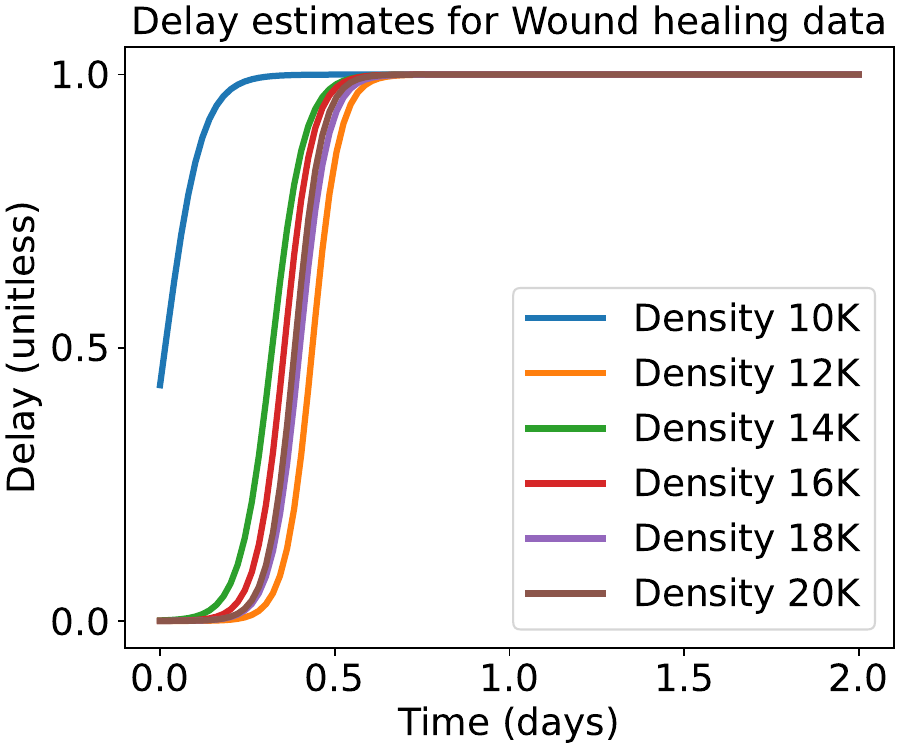}
    \caption{Wound Healing delay estimates using PDE Model 4. Estimates for the time delay, $\Tau(t)$, using PDE Model and the $\pmeanhat$ estimate for the Wound Healing datasets (Table \ref{tab:parameter_estimation_WH}).}
    \label{fig:delay_terms_WH_PDE_model_4}
\end{figure}

\begin{table}[ht!]
    \centering
    \begin{tabular}{|c|c|}
    \hline
    Number of prior samples & AIC (MSE) \\
    \hline
    $N=10^4$    &  842.4 (89.00) \\
    $N=10^5$    &  753.5 (55.05) \\
    $N=10^6$    &  720.0 (45.92) \\
    \hline
    \end{tabular}
    \caption{Slow convergence when calibrating PDE Model 4 to the Density 10 dataset. AIC scores (with MSE values in parentheses) as we vary the number of prior samples considered for the ABC algorithm when calibrating PDE Model 4 to the Density 10 dataset.}
    \label{tab:ABC_convergence_WH}
\end{table}

\section{Results using MAP}\label{sec:MAP}

\subsection{Results for the Artificial datasets}

When considering the accuracy of the $\pmap$ parameter estimator for Artificial datasets 1-4 (Supplementary Table \ref{tab:artificial_PE_map}), the PDE and ABM models perform comparably, but the $\pmaphat$ estimates are slightly more accurate than the $\pmeanhat$ estimates for Wound Healing datasets 2 and 4 while the $\pmaphat$ estimates are significantly worse for Wound Healing datasets 1 and 3 (See Table \ref{tab:artificial_PE_mean}). Using the $\pmap$ parameter estimator with either model for Artificial datsets 6-8 leads to poor estimation of the $a_0$ and $a_1$ parameters (Supplementary Table  \ref{tab:artificial_PE2_map}).

\begin{table}[ht!]
\begin{tabular}{|c|llll|}
\hline
  & \multicolumn{1}{c}{Dataset 1} & \multicolumn{1}{c}{Dataset 2} & \multicolumn{1}{c}{Dataset 3} & \multicolumn{1}{c|}{Dataset 4} \\
\hline
True & (0.031, 0.500)$^T$ & (0.031, 2.000)$^T$ & (0.156, 0.500)$^T$ & (0.156, 2.000)$^T$ \\
PDE & (0.035, 0.510)$^T$, [0.127] & (0.034, 1.904)$^T$, [0.095] & (0.133, 0.464)$^T$, [0.167] & (0.168, 1.942)$^T$, [0.079] \\
ABM & (0.018, 0.466)$^T$, [0.434] & (0.033, 2.038)$^T$, [0.056] & (0.135, 0.517)$^T$, [0.138] & (0.156, 2.034)$^T$, [0.017] \\
\hline
\end{tabular}
\caption{Parameter estimates for Datasets 1-4 when using PDE Model 1 and ABM 1  and the $\pmaphat$ parameter estimate. We report the parameter estimate value and its RED value from Equation \eqref{eq:RED} in comparison to the true parameter, $\ptrue$. \label{tab:artificial_PE_map}}
\end{table}

\FloatBarrier 
\begin{sidewaystable}[htbp]
{\small
\begin{tabular}{|c|llll|}
\hline
  & \multicolumn{1}{c}{Dataset 5} & \multicolumn{1}{c}{Dataset 6} & \multicolumn{1}{c}{Dataset 7} & \multicolumn{1}{c|}{Dataset 8} \\
\hline
True & (0.031, 0.500, 0.500)$^T$ & (0.156, 2.500, 0.000, 5.000)$^T$ & (0.156, 2.500, -1.250, 5.000)$^T$ & (0.156, 0.500, 2.500, 0.000, 5.000)$^T$ \\
PDE & (0.026, 0.500, 0.473)$^T$, [0.185] & (0.149, 2.423, -3.750, 30.000)$^T$, [6.897] & (0.156, 2.386, -8.000, 30.000)$^T$, [7.359] & (0.141, 0.500, 2.100, 0.000, 16.667)$^T$, [2.546] \\
ABM & (0.040, 0.167, 0.492)$^T$, [0.725] & (0.144, 2.464, -2.667, 20.000)$^T$, [4.738] & (0.177, 2.206, -6.250, 36.000)$^T$, [7.380] & (0.134, 0.375, 2.192, 18.000, 4.000)$^T$, [17.004] \\
\hline
\end{tabular}
}
\caption{Parameter estimates for Artificial datasets 5-8 when using  the PDE or ABM version of the true underlying model  and the $\pmaphat$ parameter estimate. For Artificial dataset 5, the true underlying model is ABM Model 2; for Artificial datasets 6 and 7, the true underlying model is ABM Model 3; and for Artificial dataset 8, the true underlying model is ABM Model 4. We report the parameter estimate value and its RED value from Equation \eqref{eq:RED} in comparison to the true parameter, $\ptrue$. \label{tab:artificial_PE2_map}}
\end{sidewaystable}
\FloatBarrier 

When computing AIC scores using the $\pmap$ estimator (Table \ref{tab:AIC_artificial_map}), the true underlying ABM model results in the lowest AIC score for one of the eight datasets, and the second-lowest AIC score for four of the eight datasets. The mean-field PDE approximation of the true underlying ABM model obtains the lowest AIC score in four of the eight datasets, and the second-lowest for one of the eight datasets. There is only one dataset  (Artificial dataset 8) where neither the true ABM nor its mean-field PDE counterpart achieves the lowest or second-lowest AIC score. When performing out-of-sample forecasting with the $\pmap$ estimator (see Supplementary Figure \ref{fig:forecasting_artificial_map}), the true ABM obtained the lowest MSE value for two  of the eight datasets (Artificial datasets 2 and 5), whereas its mean-field counterpart ABM achieved the best performance for four of the eight datasets (Artificial datasets 3, 4, 6, and 8).

\FloatBarrier 
\begin{sidewaystable}[htbp]
\begin{tabular}{|l|cccccccc|}
\hline
 & Dataset 1 & Dataset 2 & Dataset 3 & Dataset 4 & Dataset 5 & Dataset 6 & Dataset 7 & Dataset 8 \\
\hline
PDE Model 1 & \uline{\cellcolor{green!25}237.7 (3.5)} & \textbf{\cellcolor{green!25}432.1 (10.01)} & \cellcolor{green!25}314.5 (5.3) & \textbf{\cellcolor{green!25}329.5 (5.75)} & \uline{279.7 (4.39)} & 426.8 (9.72) & 605.3 (25.52) & 532.8 (17.24) \\
PDE Model 2 & 250.4 (3.71) & \uline{438.3 (10.24)} & 314.4 (5.24) & 363.5 (6.83) & \cellcolor{green!25}292.9 (4.66) & 429.1 (9.74) & 632.6 (29.25) & \textbf{507.2 (14.86)} \\
PDE Model 3 & \textbf{223.2 (3.17)} & 492.1 (13.54) & \textbf{296.6 (4.71)} & 341.9 (6.02) & 296.0 (4.69) & \textbf{\cellcolor{green!25}350.0 (6.28)} & \textbf{\cellcolor{green!25}390.9 (7.84)} & 530.4 (16.66) \\
PDE Model 4 & 248.5 (3.59) & 459.9 (11.26) & 300.9 (4.77) & 348.1 (6.15) & 287.0 (4.42) & 418.7 (9.01) & 709.6 (43.41) & \cellcolor{green!25}534.1 (16.81) \\
ABM Model 1 & \cellcolor{blue!25}306.3 (5.07) & \cellcolor{blue!25}447.1 (10.85) & \uline{\cellcolor{blue!25}300.4 (4.91)} & \uline{\cellcolor{blue!25}341.7 (6.14)} & 308.2 (5.12) & 449.4 (10.99) & 612.9 (26.59) & 519.8 (16.08) \\
ABM Model 2 & 343.4 (6.13) & 488.9 (13.45) & 319.0 (5.37) & 397.8 (8.22) & \textbf{\cellcolor{blue!25}255.8 (3.82)} & 467.8 (12.01) & 625.8 (28.2) & \uline{508.9 (14.99)} \\
ABM Model 3 & 296.6 (4.71) & 538.5 (17.41) & 359.7 (6.62) & 413.6 (8.86) & 369.8 (6.99) & \uline{\cellcolor{blue!25}403.1 (8.37)} & \uline{\cellcolor{blue!25}494.8 (13.74)} & 521.6 (15.89) \\
ABM Model 4 & 581.1 (21.67) & 583.2 (21.93) & 338.9 (5.85) & 366.3 (6.79) & 382.8 (7.42) & 479.3 (12.5) & 601.5 (24.2) & \cellcolor{blue!25}545.6 (17.89) \\
\hline
\end{tabular}
\caption{Model selection for the Artificial dataset using the $\pmap$ parameter estimator. AIC scores (with MSE values in parentheses) for each Artificial dataset. For each column, \textbf{the bolded value is the lowest AIC value}, and \uline{the underlined value is the second-lowest AIC value}, the blue cell contains the AIC score of the true model, and the green cell contains the AIC score of the PDE version of the true model.} \label{tab:AIC_artificial_map}
\end{sidewaystable}
\FloatBarrier 

\begin{figure}[ht!]
    \centering
    \includegraphics[width=0.85\linewidth]{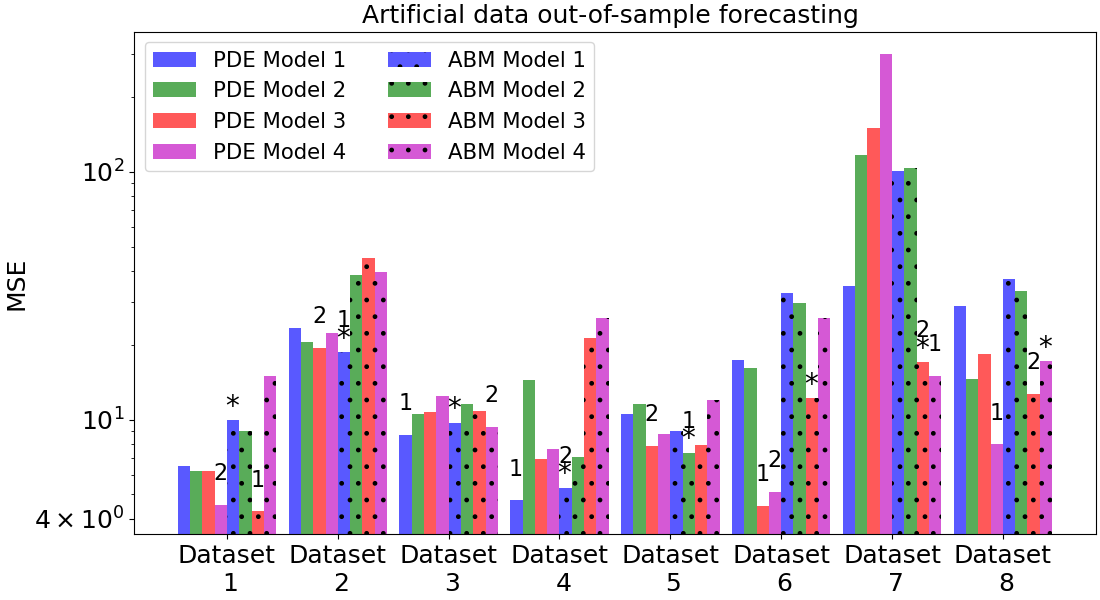}
    \caption{Out-of-sample forecasting for the Artificial datasets using the $\pmaphat$ estimate. We depict the MSE between the final Artificial data timepoint, and the prediction of each model at the final timepoint using the $\pmaphat$ parameter estimate (determined from all timepoints, excluding the final one). The true model used to generate each simulation is designated by the asterisk; the `1' and `2' designate the models with  best and second-best out-of-sample MSEs, respectively.}
    \label{fig:forecasting_artificial_map}
\end{figure}

\subsection{Results for the Wound Healing datasets}

When using the $\pmap$ estimator to compute AIC values for the Wound Healing data (Supplementary Table \ref{tab:AIC_WH_map}), we find that PDE Model 4 achieves the lowest AIC for three of the six datasets (Densities 14K, 18K, and 20K), ABM Model 4 achieves the lowest AIC for two of the six datasets (Densities 12K and 16K), and ABM Model 1 achieves the lowest AIC for 1 dataset (Density 10K).

\begin{table}[ht!]
\begin{tabular}{|l|cccccc|}
\hline
MAP & Density 10K & Density 12K & Density 14K & Density 16K & Density 18K & Density 20K \\
\hline
PDE Model 1 & 722.6 (48.12) & 685.9 (39.45) & 708.4 (44.55) & 724.3 (48.56) & 737.1 (52.03) & 800.1 (73.15) \\
PDE Model 2 & 734.3 (50.71) & 669.1 (35.64) & 719.7 (46.85) & 672.3 (36.27) & 732.9 (50.31) & 797.4 (71.31) \\
PDE Model 3 & 744.1 (52.89) & 626.4 (27.99) & 646.1 (31.13) & 691.2 (39.73) & 810.8 (75.86) & \uline{690.0 (39.47)} \\
PDE Model 4 & 883.5 (111.12) & 640.0 (29.81) & \textbf{598.2 (23.77)} & 668.9 (34.84) & \textbf{665.0 (34.11)} & \textbf{664.1 (33.95)} \\
ABM Model 1 & \textbf{705.3 (43.82)} & 693.3 (41.06) & 678.3 (37.87) & 682.1 (38.64) & 708.0 (44.45) & 769.6 (62.02) \\
ABM Model 2 & \uline{706.7 (43.68)} & 662.9 (34.47) & 668.9 (35.6) & 664.9 (34.85) & 721.0 (47.19) & 776.9 (63.84) \\
ABM Model 3 & 713.2 (44.74) & \uline{617.0 (26.6)} & 686.4 (38.71) & \uline{658.1 (33.23)} & 760.4 (57.76) & 891.0 (116.99) \\
ABM Model 4 & 716.4 (45.04) & \textbf{600.8 (24.11)} & \uline{622.4 (27.09)} & \textbf{638.0 (29.48)} & \uline{694.9 (40.1)} & 833.0 (84.61) \\
\hline
\end{tabular}
\caption{Model selection for the Wound Healing data using the $\pmap$ parameter estimator. AIC scores (with MSE values in parentheses) for each Artificial dataset. For each column, \textbf{the bolded value is the lowest AIC value}, and \uline{the underlined value is the second-lowest AIC value}.} \label{tab:AIC_WH_map}
\end{table}

\begin{figure}[ht!]
    \centering
    \includegraphics[width=0.85\linewidth]{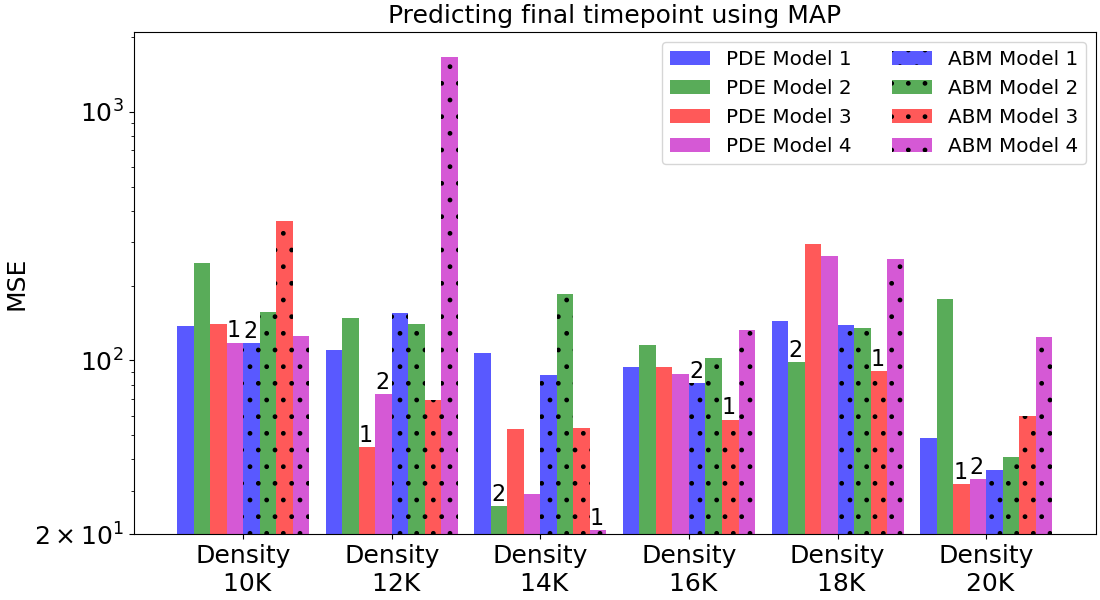}
    \caption{Out-of-sample forecasting for the Wound Healing datasets using the $\pmaphat$ estimate. We depict the MSE between the final Wound Healing data timepoint, and the prediction of each model at the final timepoint using the $\pmaphat$ parameter estimate (determined from all timepoints, excluding the final one). The `1' and `2' designate the models with  best and second-best out-of-sample MSEs, respectively.}
    \label{fig:forecasting_WH_map}
\end{figure} \newpage

\end{document}